\documentclass[aps,prx,reprint,longbibliography,floatfix]{revtex4-2}
\usepackage{amsmath,amssymb,amsthm}

\usepackage{booktabs}
\usepackage{graphicx}
\usepackage{braket}
\usepackage[colorlinks=true,allcolors=blue]{hyperref}

\newtheorem{theorem}{Theorem}
\newtheorem{lemma}[theorem]{Lemma}
\newtheorem{proposition}[theorem]{Proposition}
\newtheorem{conjecture}[theorem]{Conjecture}
\newtheorem{corollary}[theorem]{Corollary}
\theoremstyle{remark}

\newcommand{\Tr}{\operatorname{Tr}}
\newcommand{\st}{\mathrm{st}}
\newcommand{\op}{\mathrm{op}}
\newcommand{\wit}{\mathrm{wit}}
\newcommand{\sdp}{\mathrm{SDP}}
\newcommand{\Gs}{\widehat{G}_\sigma}
\newcommand{\Pp}{\Pi_\perp}
\providecommand{\openone}{\mathbb{I}}
\newcommand{\Id}{\openone}
\newcommand{\SU}{\mathrm{SU}}
\newcommand{\Eah}{\mathbb{E}_a}
\begin{document}

\title{Beyond the Stabilizer Norm: Physicality and Resource Allocation in Pauli-Limited Quantum Certification}

\author{Shreya Savadatti}
\email{ssavadat@andrew.cmu.edu}
\author{Peter Zhang}
\affiliation{Carnegie Mellon University, Pennsylvania, USA}

\begin{abstract}
Certification from restricted and uncertain Pauli measurements can be posed as a robust optimization
problem: among all physical channels consistent with the data, how small can the fidelity with a target
channel be? We first study a linear relaxation that ignores positivity and show that it is exactly
characterized by a single scalar, the stabilizer norm of the target Choi state,
$s_\wit(U)=\|C_U\|_\st-1/D$. We then ask whether this scalar continues to determine robustness once the
positivity constraint is restored. It does not. We derive a necessary and sufficient KKT
condition for the relaxation to remain tight, together with a dual obstruction certifying a strict gap,
and give a rigorous computer-assisted proof that two distinct two-qubit targets can have exactly the same stabilizer norm but
different physical robustness. The KKT condition depends on structure discarded by the stabilizer norm,
including the support and relative signs of the Choi-Pauli coefficients and the algebraic relations
among the corresponding Pauli operators. We use this
structure to derive simpler sufficient certificates for a strict gap. The dual coefficients also determine
how measurement uncertainty enters the finite-data fidelity bound. For a fixed budget, we derive exact
shot-allocation rules for a fixed certificate and
for compatible observables obtained from shared measurement settings, and we formulate the corresponding
joint certificate-design problem. The conjecture that every full-support two-qubit target has positive gain remains open.
\end{abstract}

\maketitle
\raggedbottom
\setlength{\parskip}{0pt}

\section{Introduction}
\label{sec:intro}

Suppose an experiment is intended to implement a target quantum gate \(U\). Instead of reconstructing the entire process, one may measure a finite collection of Pauli expectation values. With finite data, these measurements determine intervals rather than exact values, and many different physical channels may be consistent with the observations. A certification procedure should therefore answer a worst-case question: among all physical channels consistent with the measured data, what is the smallest possible fidelity with the target? A large lower bound then certifies the target without requiring full process tomography~\cite{Cert,Eisert}.

In the Choi representation, this question becomes a semidefinite optimization problem. The unknown Choi matrix must reproduce the measured data, satisfy trace preservation, and be positive semidefinite. The fidelity with the target is linear in this matrix. If the positivity constraint is removed, one obtains a simpler affine relaxation, or equivalently an optimized linear fidelity witness. This relaxation can only give a weaker certificate because it also allows matrices that do not represent physical quantum channels. This raises the question:

\emph{When does enforcing physicality improve a Pauli-based fidelity certificate, and what information about the target determines that improvement?}

We analyze small deviations from the ideal target measurements. Starting from the exact Pauli expectations of \(U\), we enlarge the allowed uncertainty by a small radius \(\rho\) and measure how quickly the worst-case fidelity decreases. We call this first-order decrease the robustness slope. A smaller slope means that the certified fidelity is less sensitive to measurement uncertainty.

For the positivity-relaxed Pauli problem, this slope can be solved exactly. If \(C_U\) is the normalized Choi projector of the target and \(D=d^2\), then
\begin{equation}
s_\wit(U)=\|C_U\|_\st-\frac{1}{D},
\label{eq:intro-linear}
\end{equation}
where \(\|C_U\|_\st\) is the stabilizer norm. The stabilizer norm therefore determines the robustness slope of the linear certificate. The stabilizer norm also appears in work on nonstabilizerness and Pauli-based estimation~\cite{Veitch,HC,SRE,Leone,FawziOptimal}.

Let \(s_\sdp(U)\) denote the robustness slope of the physical semidefinite problem and define the positivity gain
\[
g(U)=s_\wit(U)-s_\sdp(U)\ge0.
\]
A target has \(g(U)=0\) when the linear relaxation is already locally tight, and \(g(U)>0\) when positivity removes a worst-case direction allowed by the linear problem.

We first derive a necessary and sufficient KKT condition for \(g(U)=0\), together with a dual obstruction that certifies \(g(U)>0\). We then prove that the stabilizer norm is not sufficient to determine the physical robustness. In particular, there exist two distinct two-qubit targets \(U_0\) and \(U_1\) such that
\begin{equation}
\|C_{U_0}\|_\st=\|C_{U_1}\|_\st,
\qquad
g(U_0)=0<g(U_1).
\label{eq:intro-separation}
\end{equation}
The two targets therefore have exactly the same optimized linear robustness but different physical robustness. The separation is established by a rigorous computer-assisted proof using outward-rounded interval arithmetic, an explicit rational certificate, and analytic perturbation bounds.

The KKT condition shows which features of the target are not captured by the stabilizer norm. The norm records the total absolute Pauli weight, but not where the coefficients occur, their relative signs, or the multiplication relations among the corresponding Pauli operators. These features enter once positivity is imposed. We use them to derive simpler spectral certificates for positive gain and obtain an exact \(3\times3\) characterization in the full-support one-qubit case. For two qubits, the conjecture that every full-support target has positive gain remains open.

The dual coefficients also affect finite-data certification. A dual coefficient with large magnitude makes the fidelity certificate more sensitive to uncertainty in the corresponding measurement. For a fixed certificate and total shot budget, the optimal allocation therefore assigns
\[
N_i^\star\propto |y_i|^{2/3}.
\]
We also derive the corresponding rule when several compatible Pauli observables are obtained from the same measurement setting and formulate the joint problem of choosing a physical certificate together with its measurement allocation.

The separation proof and analytic results are established independently of the numerical studies. We separately use interval-certified calculations for finite target sets and numerical studies to examine how often the sufficient certificates succeed for sampled targets and higher dimensions. The numerical studies help assess the sufficient certificates but do not resolve the full-support two-qubit conjecture.

\section{Related work and scope}
\label{sec:related}

Several approaches to quantum-process characterization are relevant here. Full process tomography aims to reconstruct an unknown channel, while certification and verification seek a more limited guarantee about its agreement with a target~\cite{Cert,Eisert}. Mayer and Knill formulated lower bounds on process fidelity from incomplete input-state information as a semidefinite optimization problem~\cite{MayerKnill}, and worst-case fidelity bounds have also been studied for incomplete Pauli data and structured error models~\cite{Kalev,Berberich}. They show that process fidelity can be bounded without full tomography. Here we fix the Pauli measurements and their uncertainties, then compare worst-case fidelity bounds with and without the positive-semidefinite constraint.

A separate line of work designs efficient protocols specifically for quantum-process and gate verification. Liu \emph{et al.} developed efficient quantum-process verification protocols using local measurements, while Zhu and Zhang and Zeng \emph{et al.} developed frameworks for efficient unitary-gate verification under local or prepare-and-measure operations~\cite{LiuProcessVerification,ZhuGateVerification,ZengGateVerification}. More recent dimension-bounded and black-box protocols address certification when trusted state preparation and measurement cannot be assumed~\cite{Noller,Miklin}. These protocols use different assumptions about preparation and measurement. We keep the measurement family fixed and examine the improvement in the worst-case fidelity bound from complete positivity.

This approach also differs from fidelity estimation. Direct fidelity estimation samples Pauli observables according to their importance for the target~\cite{FL}, while other approaches use selected expectation values, complementary classical fidelities, or average gate fidelities~\cite{Hofmann,daSilva,Roth}. Minimax and sample-optimal estimators have been developed for fixed binary and Pauli measurement interfaces~\cite{Seshadri,FawziOptimal,Acharya}. In particular, Ref.~\cite{FawziOptimal} relates Pauli-based fidelity-estimation complexity to an \(\ell_1\) characteristic-function quantity. Here an analogous \(\ell_1\) quantity gives the exact robustness slope of the linear relaxation. We seek a lower bound that holds for every physical channel consistent with the data, rather than an estimate of a single fidelity value.

Positivity is also known to improve quantum tomography and constrained reconstruction by excluding matrices that cannot represent physical states or channels~\cite{Baldwin,Flammia2,Faist}. Here positivity constrains the worst-case fidelity minimization directly. We study its effect on the optimum of the certification problem rather than on a reconstructed estimate. The KKT condition characterizes when the linear Pauli certificate remains tight. The equal-stabilizer-norm separation then shows that the stabilizer norm alone cannot predict the improvement from positivity.

The stabilizer norm connects the linear problem to resource theories of nonstabilizerness and to the cost of Pauli-based estimation and simulation~\cite{Veitch,HC,SRE,Leone}. Once positivity is imposed, however, the support and relative signs of the Choi-Pauli coefficients and the algebraic relations among the corresponding Pauli operators become relevant. Commuting and anticommuting Pauli families are used below as tools for exposing this additional structure~\cite{Tolerant,Gross,Context,Mermin,Sarkar}.

\section{Certification framework}
\label{sec:framework}

\subsection{Choi representation and notation}

We formulate the certification problem in the Choi representation. Let $d=2^n$ be the Hilbert-space
dimension of an $n$-qubit system and let $D=d^2$. For a channel $\mathcal E$, define the normalized
Choi state
\begin{equation}
\begin{gathered}
X=(\mathcal E\otimes\mathcal I)
\big(\ket{\Phi^+}\!\bra{\Phi^+}\big),\\[2pt]
\ket{\Phi^+}=d^{-1/2}\sum_j\ket{jj}.
\end{gathered}
\end{equation}
We take register~$1$ to be the channel output and register~$2$ to be the reference. A physical
trace-preserving channel then satisfies
\begin{equation}
X\succeq0,
\qquad
T(X):=\Tr_{\rm out}(X)=\Id_d/d.
\label{eq:choi_constraints}
\end{equation}
Thus complete positivity becomes the semidefinite constraint $X\succeq0$, while trace preservation is
an affine constraint. The linear relaxation retains trace preservation but drops positivity.

With our register ordering, $T=\Tr_1$, and its adjoint is
\begin{equation}
T^*(\Lambda)=\Id_d\otimes\Lambda.
\end{equation}
The range of $T^*$ therefore lies in the reference-local operator sector. 

We use the Hilbert-Schmidt inner product
$\braket{A,B}:=\Tr(A^\dagger B)$. For a target unitary $U$, define
\begin{equation}
\begin{gathered}
\ket{\Phi_U}
=(U\otimes\Id)\ket{\Phi^+}
=(\Id\otimes U^{\mathsf T})\ket{\Phi^+},\\
C_U=\ket{\Phi_U}\!\bra{\Phi_U}.
\end{gathered}
\end{equation}
The entanglement fidelity with the target is then simply
\begin{equation}
F_e=\braket{C_U,X}.
\end{equation}
For a unitary target it is related to the average gate fidelity by
\begin{equation}
F_{\rm avg}=\frac{dF_e+1}{d+1}.
\end{equation}
The objective of the certification problem is therefore linear in the unknown Choi matrix.

We next define the Pauli coefficients used throughout the paper. Let $P_k$ denote Pauli strings on the $2n$ Choi qubits and write
\begin{equation}
c_k(M)=\frac{\Tr(P_kM)}{D}
\end{equation}
for the Pauli coefficient of an operator $M$. We call a Pauli string \emph{register-nontrivial} if it
acts nontrivially on both the output and reference registers.

The target Choi state $C_U$ is maximally entangled. Its reduced state on either register is therefore
maximally mixed, so every nonidentity Pauli operator acting on only one register has zero expectation.
Consequently, the nontrivial Pauli weight of $C_U$ lies entirely in the two-register sector, with
\begin{equation}
\sum_{\mathrm{r.n.}} c_k(C_U)^2=\frac{D-1}{D^2}.
\end{equation}
We say that $U$ has \emph{full register-nontrivial Pauli support}, or simply \emph{full support}, when
every register-nontrivial coefficient $c_k(C_U)$ is nonzero. The sign pattern of these coefficients enters the optimality conditions below.

\subsection{Uncertainty model and robustness slope}

We first define the local robustness problem around ideal target data. Let
$A_1,\ldots,A_m$ be measured observables satisfying $\|A_i\|_{\op}\le1$, and let
\begin{equation}
b_i(U)=\braket{A_i,C_U}
\end{equation}
be the expectation value predicted by the target. We allow each measured expectation to deviate from
its target value by at most a common radius $\rho$:
\begin{equation}
\big|\braket{A_i,X}-b_i(U)\big|\le\rho,
\qquad i=1,\ldots,m.
\end{equation}
The resulting worst-case fidelity is
\begin{equation}
\begin{aligned}
L_U(\rho)=\min_X\Big\{\braket{C_U,X}\ :\ &
X\succeq0,\quad T(X)=\tfrac{\Id_d}{d},\\
&\big|\braket{A_i,X}-b_i(U)\big|\le\rho
\quad\forall i\Big\}.
\end{aligned}
\label{eq:primal}
\end{equation}
This semidefinite program minimizes fidelity over physical channels consistent with the stated measurement uncertainty.

The parameter $\rho$ is used here to study the local geometry of the certification problem. It should
not yet be interpreted as a statistical confidence radius. At $\rho=0$, the constraints reproduce the
ideal target expectations exactly. When $L_U(0)=1$, those exact data are sufficient to certify unit
fidelity with the target. We then define the \emph{robustness slope}
\begin{equation}
s(U)
=
-\frac{d}{d\rho}L_U(\rho)\Big|_{\rho=0^+}.
\label{eq:slope}
\end{equation}
Thus $s(U)$ measures the first-order decrease in the guaranteed fidelity when the allowed error in each
measured expectation is increased. Smaller $s(U)$ means that the certificate is less sensitive to small
measurement uncertainty.

The right derivative in Eq.~\eqref{eq:slope} exists because the dual representation expresses
$L_U(\rho)$ as the maximum of affine functions of $\rho$. Hence $L_U$ is convex and nonincreasing in
$\rho$.

The target-centered model above is deliberately separated from the finite-data problem. In an
experiment one observes sample means $z_i$, rather than the ideal values $b_i(U)$. A confidence region
then has the form
\begin{equation}
\big|\braket{A_i,X}-z_i\big|\le r,
\end{equation}
where $r$ is a statistical radius. In general $z_i-b_i(U)\ne0$, so the observed displacement from the
target contributes to the final fidelity bound in addition to the statistical uncertainty $r$.
Section~\ref{sec:ops} derives the corresponding finite-confidence certificate.

The dual formulation makes the role of the measured observables explicit. Strong duality holds for
Eq.~\eqref{eq:primal}; see Appendix~\ref{app:duality}. A dual certificate consists of
$(y,\Lambda,S)$ satisfying
\begin{equation}
C_U
=
\sum_i y_iA_i
+
T^*(\Lambda)
+
S,
\qquad
S\succeq0.
\label{eq:dualcert}
\end{equation}
When $L_U(0)=1$, the target $C_U$ is primal optimal. For a dual certificate attaining this value,
complementary slackness gives
\begin{equation}
S\ket{\Phi_U}=0.
\end{equation}
The three terms have distinct roles. The coefficients $y_i$ weight the measured observables,
$T^*(\Lambda)$ accounts for the trace-preservation constraint, and the positive-semidefinite operator
$S$ is the contribution available only when physicality is used.

For the box uncertainty above, enlarging the radius by $\rho$ penalizes a dual certificate by
$\rho\|y\|_1$. At $\rho=0$ there may be several dual certificates that attain the target value. The
right derivative is determined by those target-optimal certificates, and the robustness slope is the
smallest $\ell_1$ cost among them. More generally, for another uncertainty set, $\|y\|_1$ is replaced
by the corresponding polar or dual norm.

The coefficients $y_i$ control sensitivity to measurement uncertainty, while the PSD slack $S$ determines whether positivity improves the bound.

\subsection{Linear relaxation and physical semidefinite certification}

We now isolate the effect of positivity. Starting from Eq.~\eqref{eq:primal}, consider the relaxation
obtained by dropping $X\succeq0$ while keeping the same measurement constraints and trace-preservation
condition. The relaxed problem uses only affine information. The physical problem uses the same
information together with complete positivity.

Since every physical Choi matrix is feasible for the affine relaxation, but not every point in the
relaxed feasible set is physical, positivity can only improve the worst-case fidelity bound. We quantify that change through the robustness slope.

For the coordinatewise $\ell_\infty$ uncertainty model, the dual cost is an $\ell_1$ norm. We now
specialize to the Pauli measurement interface used for the principal robustness results. We measure all
Pauli strings that are nonidentity on the output register. There are
\begin{equation}
d^2(d^2-1)
=
(d^2-1)^2+(d^2-1)
\end{equation}
such coordinates: $(d^2-1)^2$ two-register correlators and $d^2-1$ output-local marginals. Pauli
operators acting only on the reference register need not be measured independently because their
expectations are fixed by trace preservation.

For an operator $M$, write
\begin{equation}
\|M\|_{P,1}
=
\sum_{k\ne\Id}|c_k(M)|.
\end{equation}
Minimization over $T^*(\Lambda)=\Id_d\otimes\Lambda$ removes the reference-local coefficients, leaving
precisely the Pauli coordinates that carry independent measurement cost. The local robustness of the
linear relaxation is therefore
\begin{equation}
s_\wit(U)
=
\min_{\Lambda}
\big\|C_U-T^*(\Lambda)\big\|_{P,1},
\label{eq:swit}
\end{equation}
whereas the physical problem has the additional PSD term
\begin{equation}
s_\sdp(U)
=
\min_{\substack{\Lambda,\;S\succeq0\\
S\ket{\Phi_U}=0}}
\big\|C_U-T^*(\Lambda)-S\big\|_{P,1}.
\label{eq:ssdp}
\end{equation}
The notation ``witness'' in $s_\wit$ refers to an optimized linear fidelity witness assembled from the
measured Pauli expectations; it does not mean that a single Pauli observable is used.

The principal interface contains $12$ measured coordinates for one qubit, $240$ for two qubits, and
$4032$ for three qubits. The output-local coordinates are included even though their ideal values on
$C_U$ vanish. This is important for the later KKT analysis: at a zero Pauli coefficient, the
$\ell_1$ subgradient is not fixed to a sign and may take any value in a bounded interval. The measured
output-local coordinates provide exactly this bounded freedom. Sparse interfaces, in which some of these
coordinates are removed, are treated separately in Sec.~\ref{sec:ops}.

Equations~\eqref{eq:swit} and \eqref{eq:ssdp} differ only through the PSD contribution $S$. We therefore
define the \emph{positivity gain}
\begin{equation}
g(U)
=
s_\wit(U)-s_\sdp(U)
\ge0.
\label{eq:gain}
\end{equation}
A target with $g(U)=0$ receives no first-order improvement from positivity: the best linear witness is
already locally tight. If $g(U)>0$, the affine relaxation admits a worst-case direction that is excluded
by the physical PSD cone. Characterizing when these two cases occur is the main subject of
Secs.~\ref{sec:gain} and~\ref{sec:beyond}.

\subsection{Experimental realization}

The Choi-Pauli expectations used above can be obtained in two standard ways. In an ancilla-assisted
implementation, one prepares $\ket{\Phi^+}$, applies the unknown channel to one half of the entangled
state, and measures Pauli operators on the output and reference registers. A single local-basis setting
can provide several compatible two-register correlators and output-local marginals from the same raw
outcomes. The same Choi expectations can also be obtained through the usual prepare-and-measure
reduction, without maintaining an explicit reference system.

Throughout the finite-data analysis we assume trusted state preparation and measurement and exact trace
preservation. Let $z_i$ denote the empirical mean for Pauli coordinate $i$. Suppose that each
coordinate is estimated from $N$ independent repetitions of a setting that supports that coordinate.
Different coordinates may share the same raw shots when they are compatible. Hoeffding's inequality~\cite{Hoeffding},
followed by a union bound over the $m$ estimated coordinates, gives
\begin{equation}
r_N
=
\sqrt{\frac{2\ln(2m/\delta)}{N}}
\label{eq:hoeffding}
\end{equation}
such that, with probability at least $1-\delta$, all $m$ empirical means lie within $r_N$ of their true
expectations.

Here $m$ counts estimated Pauli coordinates, not the number of distinct measurement settings. This
distinction is important because one setting may supply several coordinates. The union bound does not
require estimates extracted from the same setting to be mutually independent; it only requires the
individual concentration bounds used for each coordinate.

Equation~\eqref{eq:hoeffding} is the confidence model used for the basic finite-shot results. Other
rigorous confidence regions for tomographic data~\cite{Flammia2,Faist} can be used instead. For example,
an ellipsoidal uncertainty region replaces the $\ell_1$ penalty in the dual problem by the appropriate
dual norm.

Statistical uncertainty is not the only source of error. If readout or calibration contributes a
separately bounded systematic bias, we denote its radius by $\bar\beta$. Unlike $r_N$, this contribution
does not decrease by taking more shots. Moreover, neither $r_N$ nor $\bar\beta$ accounts for the actual
displacement of the observed data from the target, $z_i-b_i(U)$. Section~\ref{sec:ops} combines these
three contributions into the finite-confidence acceptance rule.

\section{Exact solution of the linear relaxation}
\label{sec:limits}

Before asking when positivity improves the certificate, we solve the positivity-relaxed problem. We
begin with an unrestricted benchmark in which the measurement observable itself may be chosen freely.
We then impose the Pauli measurement restriction used in the rest of the paper. The comparison separates
a universal limitation of fidelity certification from the target dependence introduced by the Pauli
interface.

\begin{theorem}[Universal unrestricted limit]
\label{thm:global}
For every unitary target $U$ and every dimension $d\ge2$, the optimal robustness slope over observables
of unit operator norm is
\begin{equation}
s^\star=\tfrac12.
\end{equation}
It is attained by
\begin{equation}
A^\star=2C_U-\Id_D,
\end{equation}
independently of the target and the dimension.
\end{theorem}

\noindent
The observable $A^\star$ has eigenvalue $+1$ on the target Choi state and $-1$ on its orthogonal
complement. Measuring it is therefore equivalent to asking whether the Choi state lies in the target
direction or outside it. Because this measurement contains the maximally entangled projector $C_U$, it
should be viewed as a theoretical benchmark rather than as a local hardware primitive.

With unrestricted measurements, the optimal local sensitivity is therefore always $1/2$, independent
of the target and system dimension. The target dependence studied below comes from restricting the available
measurement interface. From this point onward, we consider Pauli-product measurements. The proof of
Theorem~\ref{thm:global} is given in Appendix~\ref{app:global}.

For the Pauli interface, the linear relaxation remains exactly solvable, but its optimum now depends on
the target.

\begin{theorem}[Pauli-witness baseline]
\label{thm:wit}
For every unitary target,
\begin{equation}
\begin{aligned}
s_\wit(U)
&=\|C_U\|_\st-\frac1D\\[2pt]
&=
\underbrace{\Big(1-\frac1D\Big)}_{\text{Pauli-interface floor}}
+
\underbrace{\big(\|C_U\|_\st-1\big)}_{\text{non-Clifford contribution}},
\end{aligned}
\label{eq:floorsurcharge}
\end{equation}
where
\begin{equation}
\|M\|_\st
=
\frac1D\sum_k|\Tr(P_kM)|
\end{equation}
is the stabilizer norm. The second term is nonnegative and vanishes if and only if $U$ is Clifford.
For Clifford targets the linear bound is already physically tight,
\begin{equation}
s_\sdp(U)=s_\wit(U)=\frac{D-1}{D}.
\end{equation}
\end{theorem}

\noindent
Thus the optimized linear Pauli robustness separates into the dimension-dependent floor $(D-1)/D$ and
the stabilizer-norm excess $\|C_U\|_\st-1$. Before positivity is used, all target dependence is
compressed into the single scalar $\|C_U\|_\st$.

The fact that the second term vanishes exactly for Clifford targets follows from the Parseval argument
in Lemma~\ref{lem:faith}. Tightness for Clifford targets follows from the explicit scaled KKT
certificate $\widehat G=DC_U-\Id$. The proofs are given in Appendix~\ref{app:wit}.

Figure~\ref{fig:magic} shows the exact dependence of the linear robustness on the stabilizer norm. This
linear relation provides the baseline against which the physical semidefinite problem will be compared
in the next sections.

\begin{figure*}[t]
\includegraphics[width=0.93\textwidth]{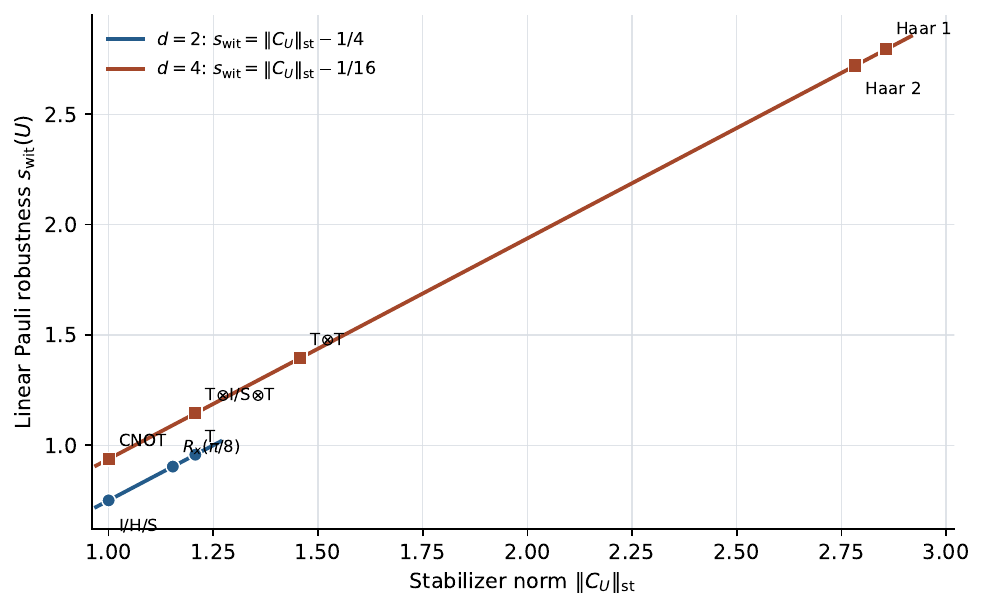}
\caption{Linear Pauli robustness as a function of the stabilizer norm. The two lines show
Eq.~\eqref{eq:floorsurcharge} for one- and two-qubit targets, and the markers indicate representative
gates. Clifford targets lie at the dimension-dependent Pauli floor. For non-Clifford targets the
additional linear sensitivity is exactly $\|C_U\|_\st-1$. The next sections show that this one-parameter
description need not survive once the positive-semidefinite constraint is imposed.}
\label{fig:magic}
\end{figure*}

\section{When is the linear relaxation tight?}
\label{sec:gain}

The previous section gives an exact solution of the linear Pauli problem. We now ask whether that solution
remains optimal when the Choi matrix is required to be positive semidefinite. Since the two problems have
the same value at the ideal data, the distinction appears only when the uncertainty region begins to
expand. The question is therefore local: do the linear and physical problems have the same initial
sensitivity to uncertainty, or does positivity improve the fidelity guarantee?

At the level of the dual problem, the answer depends on the freedom in the subgradient of the Pauli
$\ell_1$ norm. Nonzero Pauli coefficients fix the corresponding subgradient signs, whereas zero
coefficients leave bounded free parameters. The linear relaxation is tight if these free parameters can
be chosen so that the resulting subgradient is compatible with the positive-semidefinite cone at the
target. The next theorem gives the exact condition.

\begin{figure}[t]
\includegraphics[width=\columnwidth]{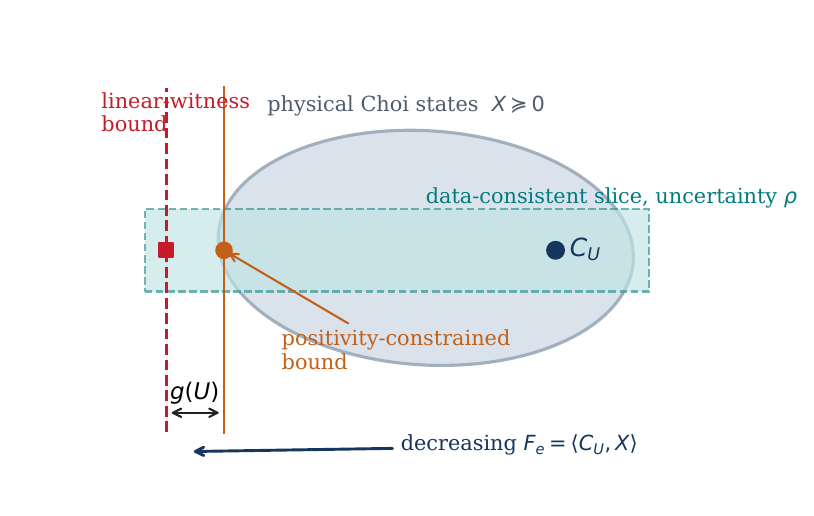}
\caption{Geometric origin of the positivity gain. The measurement constraints and trace preservation
define an affine data-consistent region. The linear relaxation minimizes the target fidelity over this
larger region without requiring the candidate Choi matrix to be positive semidefinite. Imposing
$X\succeq0$ removes nonphysical points and can therefore increase the worst-case fidelity. The
corresponding reduction in the local robustness slope is
$g(U)=s_\wit(U)-s_\sdp(U)$. The schematic is not drawn to scale.}
\label{fig:geom}
\end{figure}

Fix a minimizer $\Lambda_0$ of Eq.~\eqref{eq:swit} and define
\begin{equation}
W_0=C_U-T^*(\Lambda_0).
\end{equation}
We use the scaled Pauli subgradient
\begin{equation}
\widehat G=D\,G
\in
D\,\partial\|\cdot\|_{P,1}(W_0),
\qquad
\widehat G=\sum_{k\ne\Id}g_kP_k .
\label{eq:scaled_subgradient}
\end{equation}
The factor $D$ is introduced only to remove the normalization in the Pauli coefficients. For every
nonzero coefficient of $W_0$,
\begin{equation}
g_k=\operatorname{sign}\!\big(c_k(W_0)\big),
\end{equation}
whereas a zero coefficient allows any $g_k\in[-1,1]$.

The coefficients associated with nonzero Pauli components are therefore fixed once $W_0$ is chosen.
Only the coefficients on its zero support remain free. We refer to choosing these free coefficients as
the \emph{completion problem}. The condition $T(\widehat G)=0$ removes the reference-local freedom.
For a full-support target, the remaining free coefficients are output-local and have the form
$m\otimes\Id$. For support-deficient targets, additional register-nontrivial free coefficients may
remain; the general criterion below includes them. The relevant
question is whether this completion can be chosen so that $\widehat G$ has no positive direction in the
subspace orthogonal to the target. The precise sector decomposition is given in
Appendix~\ref{app:kkt}.

\begin{theorem}[Subgradient criterion]
\label{thm:kkt}
The positivity gain vanishes, $g(U)=0$, if and only if there is an optimal $\Lambda_0$ for
Eq.~\eqref{eq:swit} and a scaled subgradient
\[
\widehat G\in D\,\partial\|\cdot\|_{P,1}(W_0)
\]
such that
\begin{equation}
T(\widehat G)=0,
\qquad
\Pp\,\widehat G\,\Pp\preceq0,
\qquad
\Pp=\Id-C_U .
\label{eq:kktcriterion}
\end{equation}
\end{theorem}

\noindent
The two conditions in Theorem~\ref{thm:kkt} have direct interpretations. The first makes the subgradient
compatible with the trace-preserving affine space. The second requires it to be nonpositive on every
direction orthogonal to the target Choi state. Together they are the KKT condition for the physical SDP
to have the same local slope as the linear relaxation.

Cauchy interlacing gives an immediate spectral consequence. Since the compression of $\widehat G$ to the
$(D-1)$-dimensional target-orthogonal subspace is negative semidefinite, Cauchy interlacing implies that
a valid $\widehat G$ can have at most one positive eigenvalue. If such an eigenvalue is present, its
eigenvector cannot be orthogonal to $\ket{\Phi_U}$. The proof of Theorem~\ref{thm:kkt} is given in
Appendix~\ref{app:kkt}.

Theorem~\ref{thm:kkt} requires searching over the free subgradient coefficients. A dual formulation gives an equivalent way to prove that no such completion
exists.

\begin{theorem}[Dual obstruction]
\label{thm:dual}
Let $G_F$ be the part of $\widehat G$ whose coefficients are fixed by the nonzero Pauli coefficients of
$W_0$, and let $\mathcal R$ denote the admissible free Pauli labels after imposing
$T(\widehat G)=0$. Then $g(U)>0$ if and only if there exists a matrix $Q$ satisfying
\begin{equation}
Q\succeq0,
\qquad
Q\ket{\Phi_U}=0,
\qquad
\Tr Q=1,
\end{equation}
and
\begin{equation}
\braket{Q,G_F}
>
\sum_{k\in\mathcal R}
\big|\braket{Q,P_k}\big|.
\label{eq:Qcert}
\end{equation}
\end{theorem}

\noindent
The matrix $Q$ is a certificate that no admissible choice of the free coefficients can make the
target-orthogonal compression negative semidefinite. The left-hand side of
Eq.~\eqref{eq:Qcert} measures the contribution that is already fixed by the target's nonzero Pauli
coefficients. The right-hand side is the largest correction that the remaining free coefficients can
supply against the same matrix $Q$. If the fixed contribution is larger, every admissible completion
retains a positive target-orthogonal direction, and the linear relaxation cannot be tight.

The equivalence follows from Sion's minimax theorem; the proof is given in
Appendix~\ref{app:kkt}. Thus Theorem~\ref{thm:kkt} searches for a valid completion, while
Theorem~\ref{thm:dual} certifies when none exists.

Figure~\ref{fig:geom} summarizes the corresponding primal geometry. The affine relaxation contains
directions that satisfy the measurement and trace-preservation constraints but need not correspond to
physical Choi matrices. If a direction responsible for the worst linear sensitivity leaves the
positive-semidefinite cone, the physical problem cannot follow it. The physical fidelity bound then
decreases more slowly with uncertainty, giving $s_\sdp(U)<s_\wit(U)$ and hence $g(U)>0$.

\section{The stabilizer norm does not determine physical robustness}
\label{sec:beyond}

Theorem~\ref{thm:wit} shows that the stabilizer norm completely determines the linear Pauli robustness:
two targets with the same stabilizer norm necessarily have the same value of $s_\wit$. The natural
question is whether the same scalar also determines $s_\sdp$. If it did, imposing positivity could
change the value of the certificate, but it could not distinguish between targets having the same
stabilizer norm.

We construct two targets with equal stabilizer norm but different positivity gains. For one target, the KKT criterion is satisfied; for the other, it is not. Since their linear slopes are identical,
their physical slopes must differ.

\begin{figure}[t]
\includegraphics[width=\columnwidth]{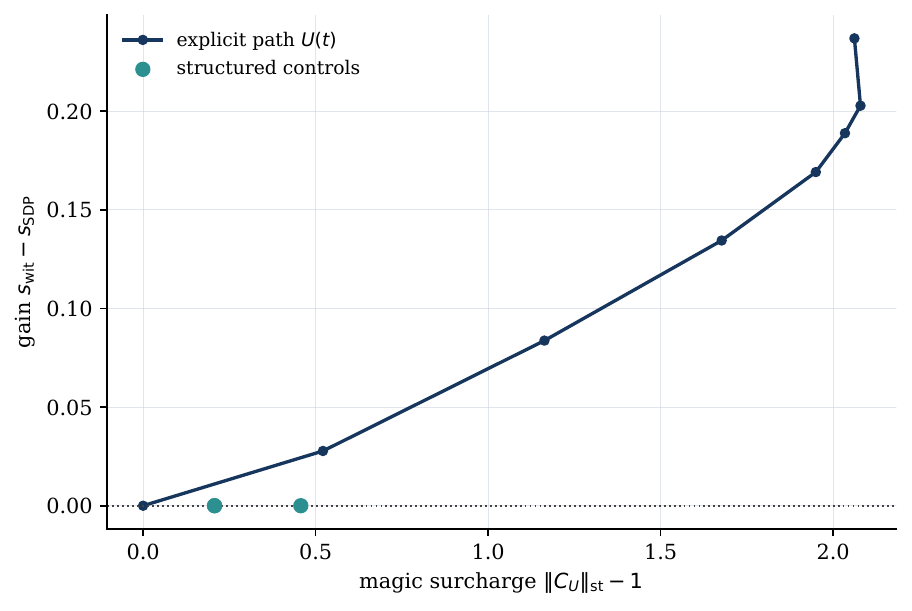}
\caption{Positivity gain for representative two-qubit targets as a function of the stabilizer norm.
The structured targets shown at zero gain demonstrate that non-Clifford stabilizer norm alone does not
force an improvement from positivity. Along the explicit path used in the separation argument, the gain
becomes positive. The certified equal-norm pair shows that two targets with the same linear robustness
can have different physical robustness.}
\label{fig:trans}
\end{figure}

We first identify an exact zero-gain target.

\begin{lemma}[Exact zero gain for $T\otimes T$]
\label{lem:tt}
\begin{equation}
g(T\otimes T)=0.
\end{equation}
\end{lemma}

\noindent
The proof constructs an admissible completion satisfying the criterion of
Theorem~\ref{thm:kkt}; see Appendix~\ref{app:main-proofs}. Thus $T\otimes T$ provides a non-Clifford
target for which positivity does not improve the local Pauli robustness.

The zero-gain target can then be matched in stabilizer norm to a target with strictly positive gain.

\begin{theorem}[Equal-stabilizer-norm separation]
\label{thm:sep}
There exist two distinct two-qubit unitary targets $U_0,U_1\in\SU(4)$ such that
\begin{equation}
\|C_{U_0}\|_\st=\|C_{U_1}\|_\st,
\qquad
g(U_0)=0<g(U_1).
\label{eq:equalnormsep}
\end{equation}
Consequently, $s_\sdp(U)$ is not determined by $\|C_U\|_\st$.
\end{theorem}

\noindent
The proof is computer-assisted but rigorous. We take
\begin{equation}
U_0=e^{-\mathrm i\pi/4}(T\otimes T)\in\SU(4).
\end{equation}
The global phase does not change the Choi projector, so
$C_{U_0}=C_{T\otimes T}$ and Lemma~\ref{lem:tt} gives $g(U_0)=0$.

To construct $U_1$, we use an explicit analytic path
\begin{equation}
U(t)=e^{-\mathrm i tH}.
\end{equation}
Outward-rounded interval arithmetic certifies an interval containing a point $t^\star$ at which
\begin{equation}
\|C_{U(t^\star)}\|_\st
=
\|C_{T\otimes T}\|_\st .
\end{equation}
Independently, a fixed rational Rayleigh certificate proves that the target-orthogonal compression has
a strictly positive direction throughout that entire interval. Analytic perturbation bounds propagate
the certificate across the interval, so in particular
\begin{equation}
g\!\left(U(t^\star)\right)>0.
\end{equation}
Taking $U_1=U(t^\star)$ proves Eq.~\eqref{eq:equalnormsep}. The generator $H$, the rational certificate,
the certified interval, the sign-stability bounds, and the perturbation estimates are given in
Appendix~\ref{app:sep}. No numerically estimated eigenvalue or operator norm is used to establish the
separation.

Theorem~\ref{thm:sep} establishes a limitation of the scalar description obtained for the linear
relaxation. The stabilizer norm determines $s_\wit(U)$ exactly, but it does not determine
$s_\sdp(U)$. Positivity therefore makes the certification problem sensitive to information about the
target that is absent from the stabilizer norm.

This separation does not identify a complete set of invariants for physical robustness. The KKT condition indicates where the additional
dependence enters. The fixed part $G_F$ depends on which register-nontrivial Choi-Pauli coefficients
are nonzero and on their signs, while the positivity condition is imposed after compression by the
target-dependent projector
\begin{equation}
\Pp(U)=\Id-C_U.
\end{equation}
Thus targets with the same total absolute Pauli weight can present different completion problems to the
positive-semidefinite cone.

\section{Pauli structure and the positivity gain}
\label{sec:geometry}

The equal-stabilizer-norm separation shows that the stabilizer norm does not contain all of the
information relevant to the physical certification problem. The KKT criterion identifies where the
additional dependence enters: the nonzero Choi-Pauli coefficients fix part of the subgradient, zero
coefficients leave a bounded completion freedom, and positivity is tested after compression onto the
subspace orthogonal to the target. We next derive spectral tests from this structure. These tests use features of the Pauli sign pattern that the stabilizer norm does not retain. The exact criterion remains Theorem~\ref{thm:kkt}.

\subsection{Product compression}

The KKT condition is exact, but checking it requires optimizing over the free completion coefficients.
A useful simplification is to restrict the target-orthogonal condition to product subspaces on which the
completion has a simple form. This leads to a lower-dimensional operator whose spectrum depends only on
the Pauli sign pattern.

For nonidentity Pauli operators $A$ and $B$ on the two Choi registers, let
$\sigma_{AB}\in\{-1,0,1\}$ denote the sign of the corresponding target coefficient, and define
\begin{equation}
\begin{gathered}
\Gs=\sum_{A\ne\Id}\sum_{B\ne\Id}\sigma_{AB}\,A\otimes B,\\[2pt]
R_\sigma(a)=\Tr_1\!\big[(\ket{a}\!\bra{a}\otimes\Id)\,\Gs\big].
\end{gathered}
\label{eq:productcompression}
\end{equation}
For each unit vector $a$, the product compression $R_\sigma(a)$ is a traceless Hermitian operator on the
second register. It records how the fixed Pauli signs appear after the first register is projected onto
$\ket a$. We write $\lambda_2(R)$ for the second-largest eigenvalue of a Hermitian operator $R$.

The second eigenvalue is the relevant quantity because the target removes one direction from the
compressed space. If $R_\sigma(a)$ has two positive eigenvalues, at least one positive direction remains
after restricting to the subspace orthogonal to the target-associated vector. Averaging over a suitable
family of $a$ then removes the remaining output-local completion.

\begin{theorem}[Product-compression certificate]
\label{thm:lp}
Let $U$ have full register-nontrivial Pauli support, with sign pattern $\sigma$.
Let $\mu$ range over probability measures on unit vectors satisfying
\begin{equation}
\int aa^\dagger\,d\mu(a)=\frac{\Id_d}{d}.
\label{eq:barycenter}
\end{equation}
If
\begin{equation}
\sup_\mu\int\lambda_2\big(R_\sigma(a)\big)\,d\mu(a)>0,
\label{eq:lp}
\end{equation}
then $g(U)>0$. If $\mu$ is restricted to a fixed finite set of directions, the optimization becomes a
finite-dimensional linear program whose value is a lower bound on the continuum supremum.
\end{theorem}

\noindent
To see why the completion disappears, fix a direction $a$. The subspace
\begin{equation}
a\otimes\big(U^{\mathsf T}\ket{a^*}\big)^\perp
\end{equation}
is orthogonal to $\ket{\Phi_U}$. On this subspace, the free part $m\otimes\Id$ of the scaled subgradient
contributes only the scalar $\braket{a|m|a}$. The barycenter condition
Eq.~\eqref{eq:barycenter} gives
\begin{equation}
\int\braket{a|m|a}\,d\mu(a)=\frac{\Tr m}{d}=0,
\end{equation}
so the completion cancels after averaging. Unlike a bound that controls the completion by its operator
norm, Eq.~\eqref{eq:lp} therefore incurs no completion penalty. The proof is given in
Appendix~\ref{app:prod}.

As a useful benchmark, the same product compression can be evaluated for a Clifford Choi stabilizer.
If $\sigma$ is its sign pattern, then
\begin{equation}
R_\sigma(a)
=d\,U^{\mathsf T}\ket{a^*}\!\bra{a^*}U^*-\Id.
\end{equation}
Its spectrum is $(d-1,-1,\ldots,-1)$ for every $a$, so $\lambda_2=-1$ pointwise and the value in
Eq.~\eqref{eq:lp} is $-1$ for every admissible $\mu$. This benchmark is consistent with $g(U)=0$,
although Clifford targets lie outside the full-support hypothesis of Theorem~\ref{thm:lp}.

\subsection{Second and third moments of the product compression}

The product-compression criterion reduces the completion problem to the spectrum of $R_\sigma(a)$. We
next ask which properties of the Pauli sign pattern affect that spectrum. The second moment gives the
first useful comparison.

\begin{lemma}[Second moment]
\label{lem:t2}
For Haar-random $a$,
\begin{equation}
\Eah\Tr R_\sigma(a)^2
=\frac1{d+1}\sum_A\Tr W_A^2,
\qquad
W_A=\sum_B\sigma_{AB}B.
\end{equation}
For a full-support sign pattern this reduces to
\begin{equation}
\Eah\Tr R_\sigma(a)^2=d(d^2-1)(d-1),
\end{equation}
independently of the sign pattern.
\end{lemma}

\noindent
The proof is given in Appendix~\ref{app:main-proofs}. For full support, the second moment is therefore
unable to distinguish one sign pattern from another. All such patterns have the same average squared
eigenvalue weight. Any difference relevant to the positivity gain must come from how this weight is
distributed across the spectrum rather than from its total magnitude.

The next moment is sensitive to the multiplication structure of the Pauli signs. For commuting Pauli
operators $A$ and $A'$, write
\begin{equation}
AA'=\omega^A(A,A')\,(A\circ A'),
\qquad
\omega^A(A,A')\in\{\pm1\},
\end{equation}
and use the analogous notation on the second register.

\begin{theorem}[Third moment and commuting triangles]
\label{thm:skew}
For Haar-random $a$,
\begin{equation}
\Eah\Tr R_\sigma(a)^3
=\frac{2d}{(d+1)(d+2)}\,\Delta_c(\sigma),
\label{eq:skew}
\end{equation}
where
\begin{equation}
\begin{aligned}
\Delta_c(\sigma)=\!\!\!\sum_{\substack{A\ne A',\,[A,A']=0\\ B\ne B',\,[B,B']=0}}\!
&\sigma_{AB}\,\sigma_{A'B'}\,
\sigma_{A\circ A',\,B\circ B'}\\
&\times\omega^A(A,A')\,\omega^B(B,B').
\end{aligned}
\label{eq:triangle}
\end{equation}
\end{theorem}

\noindent
The proof is given in Appendix~\ref{app:skew}. In the Haar average, anticommuting contributions cancel.
The surviving terms are triples whose Pauli products close within commuting pairs on both registers.
Equation~\eqref{eq:triangle} sums these multiplication triangles with the signs inherited from the target.

The second and third moments therefore retain different information. For full-support patterns, the
second moment is fixed, whereas the third moment depends on whether the Pauli signs reinforce or cancel
under commuting multiplication. The statistic $\Delta_c(\sigma)$ is thus a sign-sensitive measure of
the spectral asymmetry of the product compression. It does not by itself determine the positivity gain,
but it identifies structure that is absent from the stabilizer norm and from the second moment.

\subsection{Stabilizer benchmark for spectral skew}

A Clifford target provides a reference point because its relaxation is known to be tight. For a Clifford
sign pattern,
\begin{equation}
R=d\ket{\psi}\!\bra{\psi}-\Id,
\qquad
\ket\psi=U^{\mathsf T}\ket{a^*},
\end{equation}
with spectrum $(d-1,-1,\ldots,-1)$. Hence
\begin{equation}
\Tr R^2=d(d-1),
\qquad
\Tr R^3=d(d-1)(d-2).
\end{equation}
The corresponding pointwise normalized skew is
\begin{equation}
\kappa_{\rm pt}(a)
:=\frac{\Tr R(a)^3}{[\Tr R(a)^2]^{3/2}}
=\kappa_{\rm stab}
:=\frac{d-2}{\sqrt{d(d-1)}}.
\end{equation}
For $d=4$,
\begin{equation}
\kappa_{\rm stab}=\frac1{\sqrt3}\approx0.5774,
\end{equation}
and the CNOT support pattern attains this value exactly.

For general sign patterns we also consider the ratio formed from averaged moments,
\begin{equation}
\bar\kappa(\sigma)
:=
\frac{\Eah\Tr R_\sigma(a)^3}
{[\Eah\Tr R_\sigma(a)^2]^{3/2}}.
\label{eq:avgskew}
\end{equation}
This quantity is not the Haar average of $\kappa_{\rm pt}(a)$; the distinction matters whenever
$\Tr R_\sigma(a)^2$ depends on $a$.

The Clifford spectrum concentrates all positive spectral weight in a single direction, while the
product-compression certificate requires a positive second eigenvalue. This makes spectral skew a useful
diagnostic for how the positive part of the spectrum is distributed. Section~\ref{sec:numerics} compares
$\bar\kappa(\sigma)$ with $\Eah\lambda_2(R_\sigma(a))$ numerically.

No extremal statement is used here. In particular, we have not proved
$\bar\kappa(\sigma)\le\kappa_{\rm stab}$ for full-support patterns, nor does a pointwise spectral bound
immediately imply such an inequality for the ratio of expectations in Eq.~\eqref{eq:avgskew}. Proving an
extremal result would require additional control of the commuting-triangle statistic
$\Delta_c(\sigma)$. The Clifford patterns also lie on support-deficient boundaries rather than in the
full-support sign chambers considered below. We therefore use spectral skew only as a structural
diagnostic and test its relation to the gap numerically.

\subsection{Completion-independent spectral certificates}

The moment formulas describe features of the sign pattern, but they do not by themselves certify
$g(U)>0$. We now derive sufficient conditions that can be checked without solving the full completion
problem.

For a full-support target, the register-nontrivial part of the scaled subgradient is fixed by the target
sign pattern. Denote this fixed operator by $G_F$. The remaining output-local completion has the form
$m\otimes\Id$. Although $m$ is unknown, its operator norm is uniformly bounded. This gives a direct
spectral test.

\begin{theorem}[Eigenvalue certificate]
\label{thm:eig}
Let $\sigma$ be a full-support sign pattern and let $G_F$ be the sign-fixed part of the scaled
subgradient. Every admissible completion $m\otimes\Id$ satisfies
\begin{equation}
\|m\|_\op\le
B_d:=\sqrt{(d^2-1)(d-1)}.
\end{equation}
In particular,
\begin{equation}
B_4=\sqrt{45}\approx6.708,
\qquad
B_8=21.
\end{equation}
If
\begin{equation}
\lambda_{\max}\!\big(\Pp G_F\Pp\big)>B_d,
\label{eq:eigcert}
\end{equation}
then $g(U)>0$.
\end{theorem}

\noindent
The proof combines the completion bound with Weyl's inequality and
Theorem~\ref{thm:kkt}; see Appendix~\ref{app:prod}. Equation~\eqref{eq:eigcert} has a simple
interpretation. The left-hand side measures a positive target-orthogonal direction already present in
the sign-fixed operator. The completion can shift that spectral edge by at most $B_d$. If the positive
eigenvalue exceeds this worst-case correction, no admissible completion can remove it.

The certificate is also stable under small changes of the target that do not change the Pauli signs.
Within a \emph{sign chamber}, meaning a connected region in which none of the relevant Choi-Pauli
coefficients crosses zero, $G_F$ is constant while $\Pp(U)$ varies continuously. A strict inequality in
Eq.~\eqref{eq:eigcert} therefore persists on an open neighborhood of every certified target.

The bound $B_d$ is convenient but conservative because it treats the completion only through its
worst-case operator norm. For two qubits, anticommuting Pauli families give smaller subspaces on which
the completion can be removed rather than bounded.

\begin{theorem}[Anticommuting-clique certificate]
\label{thm:anticlique}
Let $U\in\SU(4)$ have full register-nontrivial Pauli support, let $\mathcal C$ be a set of pairwise
anticommuting nonidentity two-qubit Paulis, and let $E_{\mathcal C}$ have columns
\begin{equation}
\ket{e_A}=(A\otimes\Id)\ket{\Phi_U},
\qquad A\in\mathcal C.
\end{equation}
Define the real symmetric compression
\begin{equation}
M_{\mathcal C}(U)
=\operatorname{Re}\!\left(E_{\mathcal C}^\dagger G_F E_{\mathcal C}\right).
\end{equation}
If
\begin{equation}
\lambda_{\max}\big(M_{\mathcal C}(U)\big)>0,
\label{eq:anticlique}
\end{equation}
then $g(U)>0$.

For two qubits, the $15$ nonidentity Paulis admit six maximum pairwise-anticommuting cliques of size
five. These cliques may be chosen so that every Pauli occurs twice and every anticommuting pair occurs
once. Let $M$ denote the full $15\times15$ real compression and define
\begin{equation}
\begin{aligned}
H_{AA}&=2M_{AA},\\
H_{AC}&=M_{AC} &&\text{if }\{A,C\}=0,\\
H_{AC}&=0 &&\text{if }[A,C]=0,\ A\ne C.
\end{aligned}
\end{equation}
Then $\lambda_{\max}(H)>0$ is also sufficient for $g(U)>0$.
\end{theorem}

\noindent
The proof is given in Appendix~\ref{app:main-proofs}. In contrast with
Theorem~\ref{thm:eig}, the clique criterion contains no operator-norm allowance for the completion and
requires no optimization over its free coefficients. It is nevertheless only a sufficient condition.
The following proposition shows that it cannot replace the full KKT criterion.

\begin{proposition}[The clique criterion is not universal]
\label{prop:clique-limit}
There exists a full-support two-qubit unitary target $Q$ for which all six maximum-clique compressions
$M_{\mathcal C}(Q)$ are negative definite. Consequently, neither the size-five clique criterion nor any
of its lower-dimensional clique-subspace restrictions certifies every full-support two-qubit target.
\end{proposition}

\noindent
The proof is given in Appendix~\ref{app:main-proofs}. Failure of the clique test does not imply
$g(U)=0$; it only shows that this sufficient condition is inconclusive.

\subsection{Exact one-qubit characterization}

The certificates above are sufficient, but they do not fully characterize positive gain for two-qubit targets. For one qubit, the completion problem can instead be reduced
to an exact $3\times3$ matrix condition.

\begin{proposition}[One-qubit characterization]
\label{prop:onequbit}
Let $U\in\SU(2)$ have full register-nontrivial Pauli support and define
\begin{equation}
\begin{aligned}
O_{ij}&=\bra{\Phi_U}\sigma_i\otimes\sigma_j\ket{\Phi_U},\\
S_{ij}&=\operatorname{sign}(O_{ij}),\qquad i,j\in\{x,y,z\}.
\end{aligned}
\end{equation}
Then $O\in O(3)$ with $\det O=-1$, and
\begin{equation}
\begin{gathered}
M(U)
:=SO^{\mathsf T}+OS^{\mathsf T}
-\Big(\textstyle\sum_{ij}|O_{ij}|\Big)\Id_3,\\[2pt]
g(U)=0
\quad\Longleftrightarrow\quad
M(U)\preceq0.
\end{gathered}
\label{eq:onequbit}
\end{equation}
Equivalently, a full-support one-qubit target has positive gain exactly when
$\lambda_{\max}(M(U))>0$.

For
\begin{equation}
\begin{gathered}
U_\star
=\frac1{\sqrt{33}}
\begin{pmatrix}
4-4\mathrm i&-1\\
1&4+4\mathrm i
\end{pmatrix},\\[3pt]
O_\star
=\frac1{33}
\begin{pmatrix}
-1&32&8\\
32&-1&8\\
-8&-8&31
\end{pmatrix},
\end{gathered}
\end{equation}
one obtains
\begin{equation}
M_\star
=\frac1{33}
\begin{pmatrix}
-47&-50&8\\
-50&-47&8\\
8&8&-35
\end{pmatrix},
\end{equation}
whose spectrum is exactly
\begin{equation}
\left\{-3,-1,\frac1{11}\right\}.
\end{equation}
Hence $g(U_\star)>0$.
\end{proposition}

\noindent
The proof is given in Appendix~\ref{app:main-proofs}. Proposition~\ref{prop:onequbit} gives a complete
criterion for full-support one-qubit targets. It does not cover support-deficient targets, because zeros
among the register-nontrivial coefficients introduce additional free subgradient coefficients. The
explicit target $U_\star$ also shows that positive gain is not a specifically two-qubit phenomenon. No
corresponding exact matrix criterion is known for $d=4$.

\subsection{The full-support two-qubit conjecture}

The sufficient certificates above detect positive gain for broad classes of full-support targets, but
they do not settle the full two-qubit problem. We state the remaining question as a conjecture.

\begin{conjecture}[Full-support dichotomy]
\label{conj:sd}
Every full-support two-qubit target has $g(U)>0$.
\end{conjecture}

The following theorem shows that almost every two-qubit unitary has full support.

\begin{theorem}[Conditional genericity]
\label{thm:cond}
The set of two-qubit unitaries with at least one vanishing register-nontrivial Choi-Pauli coefficient has
Haar measure zero. Consequently, if Conjecture~\ref{conj:sd} holds, then $g(U)>0$ for Haar-almost every
two-qubit unitary.
\end{theorem}

\noindent
The measure-zero statement is unconditional; only the conclusion about positive gain depends on
Conjecture~\ref{conj:sd}. To see the first part, each register-nontrivial coefficient $c_k(U)$ is a real
polynomial in the real and imaginary parts of the entries of $U$, restricted to $\SU(4)$, and is
therefore real analytic. No such coefficient is identically zero. Appendix~\ref{app:nonzero} constructs
an explicit unitary with $c_k\ne0$ for one representative Pauli label, and two-sided Clifford actions
permute the register-nontrivial labels transitively up to sign. The zero set of each coefficient is
therefore a proper real-analytic subset of $\SU(4)$ and has Haar measure zero~\cite{KrantzParks}. Since there are finitely
many relevant coefficients,
\begin{equation}
V=\bigcup_k\{U:c_k(U)=0\}
\end{equation}
also has Haar measure zero. If Conjecture~\ref{conj:sd} is true, every zero-gain target must lie in $V$,
which gives the conditional Haar-almost-everywhere statement.

The theorem does not establish generic positive gain without the conjecture. The moment calculation
shows that commuting-triangle structure distinguishes full-support sign patterns at third order, but a
third moment does not by itself control the positive second eigenvalue required by the
product-compression certificate. We examine this conjecture numerically in the next section without claiming a general proof.

\section{Numerical tests of the relaxation gap}
\label{sec:numerics}

We compare direct semidefinite calculations with the KKT and spectral tests, then examine Pauli-sign statistics.

We distinguish throughout between certified calculations and numerical evidence. Results obtained from
exact constructions or outward-rounded interval arithmetic are rigorous. Solver optima, binary64
evaluations, and frequencies observed over finite Haar samples are used only as numerical evidence and
do not establish universal statements. Appendix~\ref{app:numprot} gives the sampling procedures, random
seeds, software versions, solver residuals, interval protocols, and standard errors used below.

\begin{figure}[t]
\includegraphics[width=\columnwidth]{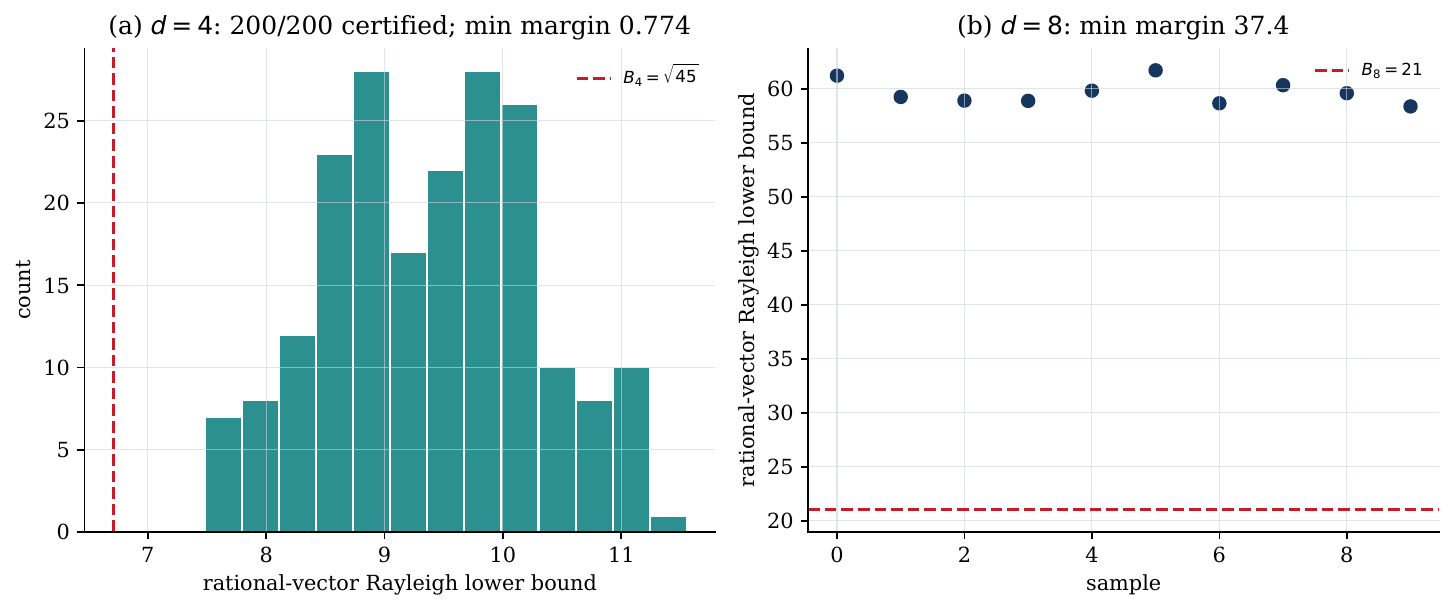}
\caption{Margins for the eigenvalue certificate of Theorem~\ref{thm:eig}. For the sampled $d=4$
targets, every certified lower bound lies above $B_4=\sqrt{45}$. The interval-certified $d=8$ targets
also lie above $B_8=21$, with larger relative margins. The finite samples do not establish an
asymptotic scaling with dimension.}
\label{fig:eig}
\end{figure}

\subsection{Consistency of the exact formulations}

As a first check, we compared the KKT characterization of Theorem~\ref{thm:kkt} with direct solution of
the physical semidefinite program. The two calculations agreed on the classification $g=0$ versus
$g>0$ for all five tested targets. This comparison checks the implementation of the KKT formulation
against the original optimization problem; it is not needed for the proof of
Theorem~\ref{thm:kkt}.

The separation is established in Theorem~\ref{thm:sep}; the numerical results below illustrate it but are not part of the proof.

\subsection{Certified gain in sampled targets}

We next tested whether the positive-gain behavior found in the constructed separation also appears for
Haar-random targets. For $200$ two-qubit unitaries drawn from Haar measure, the eigenvalue criterion
of Theorem~\ref{thm:eig} certified $g(U)>0$ for every target. These are rigorous certificates rather
than classifications based on eigensolver output. For each target, a stored trial vector gives a
Rayleigh-quotient lower bound on the relevant eigenvalue, and the bound was verified using
outward-rounded interval arithmetic. The interval verification also accounts for the transfer to the exact polar unitary
and verifies that the Pauli sign pattern is stable over the enclosure. The smallest certified margin
above
\begin{equation}
B_4=\sqrt{45}
\end{equation}
was $0.773$.

A strict certificate inequality persists under sufficiently small perturbations of the target. Each of
the $200$ certified targets therefore belongs to an open neighborhood with positive gain. This certifies a neighborhood of each target, not an entire sign chamber; several sampled targets may also lie in the same chamber.

We also applied the same test in higher dimension. All ten interval-certified $d=8$ targets exceeded
\begin{equation}
B_8=21,
\end{equation}
with a smallest certified margin of $37.38$. These margins are larger relative to the completion budget
than those in the $d=4$ sample, as shown in Fig.~\ref{fig:eig}. A separate numerical calculation on
$200$ stored three-qubit Haar targets also satisfied the sufficient inequality for every target, with a
minimum observed margin of $34.498$. The latter calculation is numerical rather than interval-certified.
The larger margins suggest studying dimension dependence, although these samples do not establish a scaling law.

The one-qubit case behaves differently. Proposition~\ref{prop:onequbit} gives an exact criterion for
every full-support one-qubit target, so no sufficient certificate is needed. Among
$2\times10^5$ Haar-random one-qubit targets,
\begin{equation}
M(U)\not\preceq0
\end{equation}
for $1.504\%$ of the sample. The binomial standard error is $0.027$ percentage points, giving the
approximate $95\%$ interval
\begin{equation}
1.504\%\pm0.053\%.
\end{equation}
Every sampled target had all relevant coefficient magnitudes above $10^{-12}$. By the same real-analytic
zero-set argument used in Theorem~\ref{thm:cond}, full support holds Haar-almost surely in the one-qubit
setting. As an independent numerical check of the
explicit example in Proposition~\ref{prop:onequbit}, direct semidefinite optimization gives
\begin{equation}
g(U_\star)\approx7.11\times10^{-4}>0.
\end{equation}
This numerical value is not used in the proof of the proposition.

\subsection{Support structure and spectral skew}

The structural results of Sec.~\ref{sec:geometry} suggest that the physical problem depends on more than
the total absolute Pauli weight. We therefore compared the support and spectral statistics of structured
two-qubit controls with those of sampled full-support targets.

The structured controls used in the numerical comparison are strongly support-deficient. Among the
$225$ register-nontrivial two-qubit coefficients, CNOT has support size $15$,
$T\otimes\Id$ and $S\otimes T$ have support size $23$, and $T\otimes T$ has support size $35$.
These controls have zero gain in the numerical comparison. The Haar targets are numerically full support.
Support deficiency alone, however, is not a general explanation for zero gain.
The one-qubit experiment above gives a clear counterexample to such an interpretation: most
full-support one-qubit Haar targets satisfy the zero-gain side of the exact criterion.

We also compared the averaged spectral skew $\bar\kappa$ with the product-compression statistic
$\Eah\lambda_2$. The structured controls have larger $\bar\kappa$ and smaller
$\Eah\lambda_2$ than the sampled Haar targets. Across $12$ targets, consisting of CNOT,
$T\otimes T$, $S\otimes T$, and nine Haar targets, each evaluated using $400$ Haar-random directions, the two quantities
have Spearman correlation
\begin{equation}
\rho_{\rm Spearman}=-0.942.
\end{equation}
The standard error of the estimated $\Eah\lambda_2$ is at most $0.052$ for these calculations.
The structured controls have
\begin{equation}
0.529\le\bar\kappa\le0.577,
\end{equation}
whereas the Haar targets have
\begin{equation}
0.303\le\bar\kappa\le0.372.
\end{equation}
Over the same comparison, $\Eah\lambda_2$ moves from $-1$ for the stabilizer-consistent reference
toward values that straddle zero.

This correlation is exploratory. It is consistent with the picture developed in
Sec.~\ref{sec:geometry}, in which concentration of spectral weight into a single positive direction is
associated with large positive skew, while a positive second eigenvalue is relevant to the
product-compression certificate. The calculation does not establish a functional relation between
$\bar\kappa$ and the positivity gain.

\subsection{Reach and limitations of the sufficient certificates}

The anticommuting-clique certificate provides a less conservative test than the operator-norm bound
because it eliminates the completion on the chosen subspace. We evaluated
Theorem~\ref{thm:anticlique} on an additional fixed-seed sample of $10{,}000$ two-qubit Haar targets.
All $10{,}000$ were numerically full support; the smallest observed magnitude of a relevant Pauli
coefficient was
\begin{equation}
3.55\times10^{-8}.
\end{equation}

Every target had at least one size-five anticommuting clique with positive compressed eigenvalue. The
smallest value observed over the sample was
\begin{equation}
\min_U\max_{\mathcal C}
\lambda_{\max}\!\big(M_{\mathcal C}(U)\big)
=1.00386.
\end{equation}
The aggregate matrix $H$ was also positive on every target, with
\begin{equation}
\min_U\lambda_{\max}(H)=0.19638.
\end{equation}
For comparison, the eigenvalue-budget test of Theorem~\ref{thm:eig} passed $9{,}999$ of the
$10{,}000$ targets. The remaining target also numerically satisfied the clique criterion. The combinatorial
structure and maximum size of anticommuting Pauli families are reviewed in
Ref.~\cite{Sarkar}; the completion-cancellation argument used for certification is derived in this work.

To examine how much of the clique structure is needed empirically, we performed a separate ten-seed
ablation in which directions were removed from the anticommuting subspaces. Three mutually
anticommuting directions gave a positive test for every target in that sample, whereas two-direction tests failed on
four targets. These are finite-sample observations, not sufficient conditions for arbitrary
full-support targets. Per-seed counts and dispersions are reported in Appendix~\ref{app:numprot}.

\begin{figure*}[t]
\centering
\includegraphics[width=0.96\textwidth]{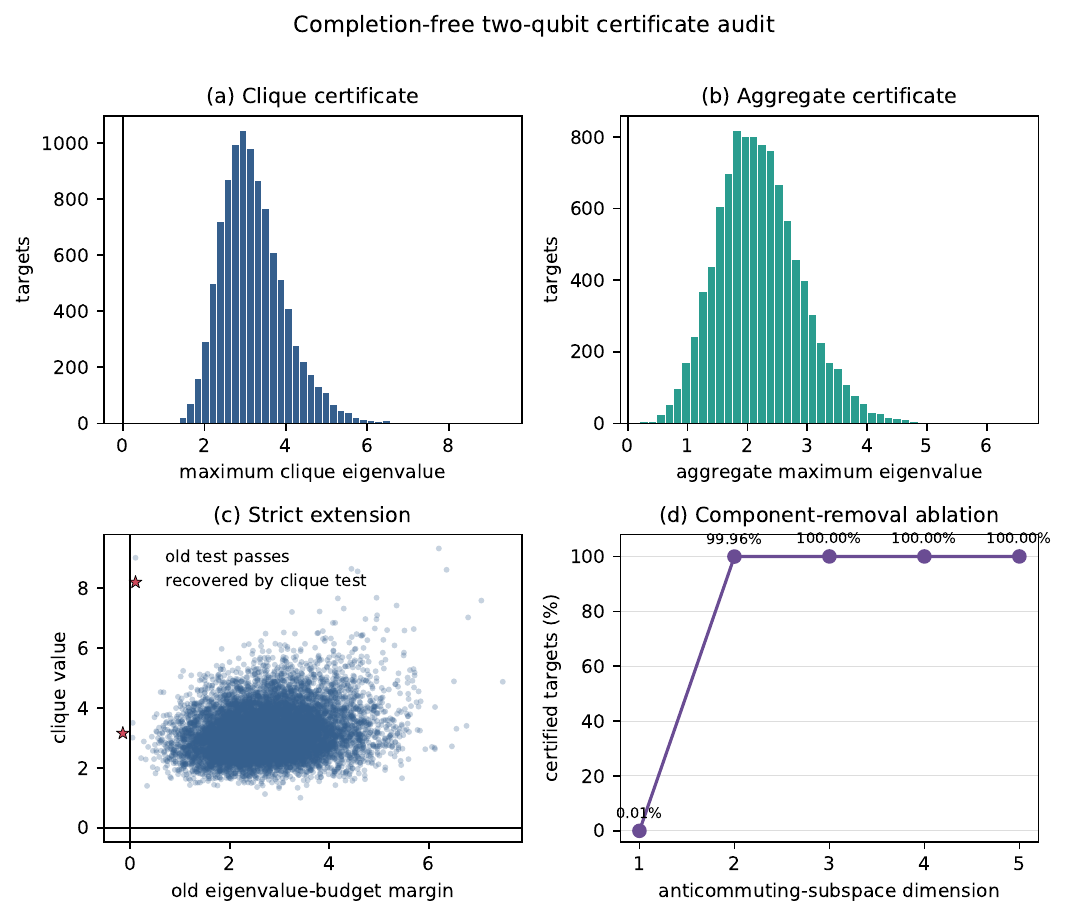}
\caption{Numerical performance of the completion-free certificates on $10{,}000$ fixed-seed
two-qubit Haar targets. (a) The largest eigenvalue among the six size-five clique compressions is
positive for every sampled target. (b) The aggregate matrix $H$ is also positive throughout the
sample. (c) The clique certificate detects the single target not certified by the operator-norm budget
test. (d) In a separate ten-seed ablation, Pauli directions are removed from the clique subspaces;
error bars show the range across ten blocks of $1{,}000$ targets.}
\label{fig:cliques}
\end{figure*}

The success of the clique test on this sample does not make it a universal criterion. The targeted
construction of Proposition~\ref{prop:clique-limit} gives a full-support target for which all six
maximum-clique compressions are negative definite. The weakest rigorous lower bound on the eigenvalues
of their negatives exceeds $0.1296$, while the polar-transfer allowance is below $10^{-9}$. Thus the
failure of the clique test is not a numerical boundary effect. It proves that the clique criterion alone
cannot establish Conjecture~\ref{conj:sd}.

This target is not a counterexample to the conjecture. The clique theorem is a sufficient condition for
positive gain, so failure of the test does not imply $g(U)=0$. To examine the remaining completion
freedom directly, we solved the full KKT completion problem for $240$ perturbations of
$T\otimes T$ at six perturbation scales down to $10^{-6}$, together with $60$ Haar controls. No
full-support zero-gain target was found. We also initialized the full completion optimization from an
adversarial target that defeats both of the simpler sufficient tests; the optimized completion value
remained positive.

The calculations use both fixed-target completion solves and a nonconvex search over unitaries. They do not certify a global minimum over $U$ and therefore do not prove Conjecture~\ref{conj:sd}.
Failure of a sufficient test therefore does not provide a full-support zero-gain example. The full KKT completion problem is needed to assess those cases.

\section{Finite-data certification and measurement allocation}
\label{sec:ops}

The robustness slope measures sensitivity to uncertainty in the Pauli expectations. Experimentally, part of that uncertainty depends on the number of shots. Increasing
the number of shots reduces statistical uncertainty, and the dual coefficients determine how uncertainty
in each measured coordinate enters the fidelity bound. We use the dual coefficients both to certify finite data and to allocate a fixed shot budget.

\subsection{Finite-confidence certificates}

We first consider a measurement family for which the ideal target data are target-complete, so that
$L_U(0)=1$. Let $(y,\Lambda,S)$ be a dual certificate attaining this value at the target data, and write
\begin{equation}
\|y\|_1=s.
\end{equation}
The certificate remains dual feasible when the target-centered uncertainty radius $\rho$ is increased,
because the feasibility condition in Eq.~\eqref{eq:dualcert} does not depend on $\rho$. Evaluating the
same certificate at any $\rho\ge0$ therefore gives
\begin{equation}
L(\rho)\ge1-s\rho,
\qquad \rho\ge0.
\label{eq:affine}
\end{equation}
Equation~\eqref{eq:affine} is a global affine lower bound for the target-centered uncertainty model,
not merely the tangent defined by the local derivative at $\rho=0$.

For experimental data, let $z_i$ denote the empirical expectation values and let $r_N$ be the
simultaneous statistical radius from Eq.~\eqref{eq:hoeffding}. On the corresponding joint confidence
event,
\begin{equation}
F_e
\ge
1+y^{\mathsf T}(z-b)-s r_N.
\label{eq:empirical-affine}
\end{equation}
The two correction terms have different meanings. The quantity
\begin{equation}
y^{\mathsf T}(z-b)
\end{equation}
depends on the displacement actually observed between the experimental data and the ideal target data.
The term $sr_N$ accounts for the remaining statistical uncertainty in those observations. The
certificate's $\ell_1$ cost therefore multiplies the finite-sample confidence radius. When the
certificate also minimizes this cost among target-optimal certificates, $s$ equals the target-centered
robustness slope.

The individual dual coefficients determine the effect of each measurement uncertainty on the bound. Uncertainty in a coordinate with a large $|y_i|$ has a larger effect on the certified
fidelity than the same uncertainty in a coordinate with a small $|y_i|$. Before using those weights to
allocate measurements, however, we first ask whether the chosen measurement family contains enough
information to certify the target.

\subsection{Information contained in a measurement family}

A measurement family need not reconstruct an entire channel in order to give a useful fidelity bound.
At one extreme, the measured data may be compatible with a physical channel orthogonal to the target.
At the other, they may identify the target uniquely. These possibilities can be distinguished directly
from the exact-data feasible set.

For a specified target $C_U$, observables $A_1,\ldots,A_m$, and ideal values
\begin{equation}
b_i=\braket{A_i,C_U},
\end{equation}
consider the physical channels that reproduce those data exactly.

\begin{proposition}[Exact-data criterion for a specified family]
\label{prop:family}
For a fixed target $C_U$, observables $A_1,\ldots,A_m$, and ideal data
$b_i=\braket{A_i,C_U}$, the following are equivalent:
\begin{enumerate}
\item the exact-data certification problem satisfies $L_U(0)=1$;
\item $C_U$ is the unique physical trace-preserving Choi state consistent with $b$;
\item there is no nonzero Hermitian perturbation $H$ satisfying
\begin{equation}
\begin{gathered}
T(H)=0,
\qquad
\braket{A_i,H}=0
\quad (i=1,\ldots,m),\\
C_U+H\succeq0.
\end{gathered}
\end{equation}
\end{enumerate}
Moreover, any real vector $y$ and Hermitian $\Lambda$ satisfying
\begin{equation}
C_U-\sum_i y_iA_i-T^*(\Lambda)\succeq0,
\qquad
b^\top y+\Tr(\Lambda)/d=1
\end{equation}
give an exact dual certificate for these conditions. Conversely, such a certificate exists whenever
the dual supremum is attained.
\end{proposition}

\noindent
The proof is in Appendix~\ref{app:main-proofs}. Condition~3 has the following interpretation. A nonzero $H$ satisfying the first two constraints changes neither trace
preservation nor any measured expectation value. If $C_U+H$ remains positive semidefinite, the same
exact data are compatible with a second physical channel. The target is uniquely identified precisely
when no such physical perturbation exists.

Uniqueness is stronger than what is required for a nonzero fidelity guarantee. To distinguish these
levels of information, define the exact-data fiber
\begin{equation}
\mathcal F_{\mathcal A}(b)
=
\left\{
X\succeq0:
T(X)=\Id_d/d,\;
\mathcal A(X)=b
\right\}.
\end{equation}

\begin{theorem}[Measurement-family hierarchy]
\label{thm:family-hierarchy}
Let
\begin{equation}
L_{\mathcal A}(0)
=
\min_{X\in\mathcal F_{\mathcal A}(b)}
\braket{C_U,X},
\end{equation}
with ideal target data $b=\mathcal A(C_U)$. Exactly one of the following holds:
\begin{enumerate}
\item
$L_{\mathcal A}(0)=0$ if and only if $\mathcal F_{\mathcal A}(b)$ contains a physical Choi state
orthogonal to $C_U$;

\item
$0<L_{\mathcal A}(0)<1$ if and only if $\mathcal F_{\mathcal A}(b)$ contains more than one physical
Choi state but none orthogonal to $C_U$;

\item
$L_{\mathcal A}(0)=1$ if and only if
$\mathcal F_{\mathcal A}(b)=\{C_U\}$.
\end{enumerate}

Define also the real trace-preserving tangent space
\begin{align}
\mathcal V_{\rm TP}
&=
\{H=H^\dagger:T(H)=0\},\\
\dim_{\mathbb R}\mathcal V_{\rm TP}
&=
d^4-d^2.
\end{align}
The measurement family $\mathcal A$ is channel-tomography-equivalent, meaning that it is injective on
the set of physical trace-preserving Choi states, if and only if
\begin{equation}
\ker\mathcal A\cap\mathcal V_{\rm TP}=\{0\}.
\label{eq:tomography-rank}
\end{equation}
Equivalently, the restriction of $\mathcal A$ to $\mathcal V_{\rm TP}$ has real rank
$d^4-d^2$.
\end{theorem}

\noindent
The proof is given in Appendix~\ref{app:main-proofs}. The three cases correspond to no positive fidelity
guarantee, partial fidelity certification, and unique target identification. A measurement family may fail even to exclude a channel orthogonal to
the target, in which case no positive fidelity can be guaranteed from the exact data. It may exclude all
orthogonal channels while leaving several physical channels compatible with the observations, giving a
strictly positive but subunit fidelity guarantee. Finally, it may identify the target itself, in which
case the exact-data fidelity bound is one.

We call the last property \emph{target completeness}. It is a target-specific notion and is weaker than
channel-tomography equivalence: a family may uniquely identify the particular target $C_U$ without
being injective on every physical channel. Target completeness is closely related to strict
completeness for a specified rank-one target in the bounded-rank tomography framework of Baldwin,
Deutsch, and Kalev~\cite{Baldwin}. The hierarchy above additionally distinguishes partial
fidelity certification from unique target identification and from full family-level channel
tomography.

\subsection{Optimal allocation for a fixed certificate}

We next optimize the shot budget for a fixed dual certificate. Suppose a dual certificate $y$ has already been chosen.
Using the same number of shots for every active observable is generally inefficient because the
coordinates do not contribute equally to the fidelity bound. The statistical penalty has the form
\begin{equation}
\sum_i |y_i|r_i,
\end{equation}
so measurements with larger $|y_i|$ should be estimated more accurately.

Let
\begin{equation}
\mathcal A_y=\{i:y_i\ne0\},
\qquad
k=|\mathcal A_y|.
\end{equation}
Assign $N_i>0$ shots to each active coordinate. If the total failure probability $\delta$ is divided
equally among the $k$ active coordinates, Hoeffding's inequality~\cite{Hoeffding} gives
\begin{equation}
r_i=\frac{\kappa}{\sqrt{N_i}},
\qquad
\kappa=\sqrt{2\ln(2k/\delta)}.
\end{equation}
For a fixed certificate and a fixed total budget, the allocation problem is therefore
\begin{equation}
\min_{\{N_i\}}
\sum_{i\in\mathcal A_y}
\frac{|y_i|}{\sqrt{N_i}}
\qquad
\text{subject to}
\qquad
\sum_{i\in\mathcal A_y}N_i=B.
\end{equation}

\begin{theorem}[Optimal shot allocation for a fixed certificate]
\label{thm:shot-allocation}
Fix a nonzero coefficient vector $y$ from a dual certificate attaining value $1$ at the ideal target
data, and let $B>0$ be a continuous shot budget. Define
\begin{equation}
\mathcal C_{2/3}(y)
:=
\left(
\sum_{i\in\mathcal A_y}|y_i|^{2/3}
\right)^{3/2}.
\end{equation}
Let $\mathcal N_B$ denote the positive allocations satisfying
\begin{equation}
\sum_{i\in\mathcal A_y}N_i=B.
\end{equation}
Then
\begin{equation}
\min_{N\in\mathcal N_B}
\sum_{i\in\mathcal A_y}
\frac{|y_i|}{\sqrt{N_i}}
=
\frac{\mathcal C_{2/3}(y)}{\sqrt B},
\label{eq:allocation-value}
\end{equation}
and the unique optimum on the active coordinates is
\begin{equation}
N_i^\star
=
B\,
\frac{|y_i|^{2/3}}
{\sum_{j\in\mathcal A_y}|y_j|^{2/3}}.
\label{eq:allocation-rule}
\end{equation}
Consequently, if coordinate $i$ also has a bounded systematic allowance $\beta_i$, then on the joint
confidence event
\begin{equation}
F_e
\ge
1+y^{\mathsf T}(z-b)
-\frac{\kappa\mathcal C_{2/3}(y)}{\sqrt B}
-\sum_i|y_i|\beta_i.
\label{eq:allocated-acceptance}
\end{equation}
\end{theorem}

\noindent
The $2/3$ exponent follows from the competition between certificate sensitivity and statistical
concentration. Increasing $N_i$ reduces the uncertainty only as $N_i^{-1/2}$, while the effect of that
uncertainty on the fidelity bound is proportional to $|y_i|$. The optimum therefore assigns more shots
to heavily weighted coordinates, but less aggressively than an allocation proportional to $|y_i|$.

In many experiments, several compatible Pauli observables are obtained from the same measurement
setting. The relevant resource is then the number of shots assigned to the setting rather than to each
coordinate separately.

\begin{corollary}[Allocation across shared measurement settings]
\label{cor:setting-allocation}
Suppose the active coordinates are partitioned into settings
\begin{equation}
\mathcal G_1,\ldots,\mathcal G_q,
\end{equation}
and every coordinate assigned to setting $s$ is estimated from the same $N_s$ shots. Define
\begin{equation}
w_s(y)
=
\sum_{i\in\mathcal G_s}|y_i|.
\end{equation}
Under the same simultaneous Hoeffding event, the statistical penalty is
\begin{equation}
\kappa\sum_s\frac{w_s(y)}{\sqrt{N_s}}.
\end{equation}
Let $\mathcal M_B$ denote the positive setting allocations satisfying
\begin{equation}
\sum_sN_s=B.
\end{equation}
For settings with $w_s(y)>0$, the unique continuous-budget optimum is
\begin{equation}
N_s^\star
=
B\,
\frac{w_s(y)^{2/3}}
{\sum_t w_t(y)^{2/3}}.
\end{equation}
Defining
\begin{equation}
\mathcal C_{\mathcal G}(y)
=
\left(
\sum_s w_s(y)^{2/3}
\right)^{3/2},
\end{equation}
the minimum statistical penalty is
\begin{equation}
\min_{N\in\mathcal M_B}
\sum_s\frac{w_s(y)}{\sqrt{N_s}}
=
\frac{\mathcal C_{\mathcal G}(y)}{\sqrt B}.
\label{eq:grouped-cost}
\end{equation}
The simultaneous union bound does not require statistical independence between coordinates extracted
from the same measurement setting.
\end{corollary}

\noindent
The proof of Theorem~\ref{thm:shot-allocation} and
Corollary~\ref{cor:setting-allocation} is given in Appendix~\ref{app:shot-allocation}. These results
optimize the measurement allocation for a fixed dual certificate. They do not optimize the certificate
itself or the choice of measurement settings.

The quantity $\mathcal C_{2/3}$ is positively homogeneous but is not a norm. The stated optimum uses
the common Hoeffding factor obtained by assigning failure
probability $\delta/k$ to each active coordinate; optimizing the confidence allocation
$\{\delta_i\}$ is a separate problem. Physical shot counts are integer-valued. For an integer budget
$B\ge k$, the exact discrete optimum is obtained by assigning one shot to each active coordinate and
then repeatedly placing the next shot on the coordinate that gives the largest decrease in
\begin{equation}
\frac{|y_i|}{\sqrt{N_i}}.
\end{equation}
The continuous solution in Eq.~\eqref{eq:allocation-rule} is the corresponding relaxation and gives the
large-budget allocation.

Finally, Corollary~\ref{cor:setting-allocation} assumes that the active coordinates form a fixed
partition into measurement settings. More general experiments may estimate the same coordinate from
several overlapping settings or exploit covariance between jointly measured observables. Such designs
require a more general optimization model and are not covered by the allocation theorem above.

\subsection{Joint optimization of the certificate and measurement budget}

The allocation results above assume that the dual certificate $y$ has already been fixed. The physical
problem allows a larger set of dual certificates than the linear relaxation because the positive
semidefinite slack $S$ may also be used. It is therefore natural to optimize the certificate and the
allocation of measurement shots together. Since several Pauli coordinates may be obtained from the same
measurement setting, the relevant experimental resource is the number of shots assigned to each raw
setting rather than the number of reported coordinates.

Let $\mathcal Y_{\rm phys}(U)$ denote the set of coefficient vectors $y$ for which there exist
$\Lambda$ and $S\succeq0$ satisfying
\begin{equation}
S\ket{\Phi_U}=0
\end{equation}
and the dual feasibility condition in Eq.~\eqref{eq:dualcert}. Let
$\mathcal Y_{\rm wit}(U)$ denote the corresponding set when $S=0$. For a fixed partition
$\mathcal G$ of the measured coordinates into compatible settings, define
\begin{align}
\Gamma_{\rm phys}(U;\mathcal G)
&=
\inf_{y\in\mathcal Y_{\rm phys}(U)}
\mathcal C_{\mathcal G}(y),
\label{eq:joint-setting-cost}\\
\Gamma_{\rm wit}(U;\mathcal G)
&=
\inf_{y\in\mathcal Y_{\rm wit}(U)}
\mathcal C_{\mathcal G}(y).
\end{align}
Thus $\Gamma_{\rm phys}$ is the smallest setting-weighted statistical cost attainable by any valid
physical dual certificate for the target. Since every linear-witness certificate is also feasible in
the physical dual,
\begin{equation}
\mathcal Y_{\rm wit}(U)
\subseteq
\mathcal Y_{\rm phys}(U),
\end{equation}
and therefore
\begin{equation}
\Gamma_{\rm phys}(U;\mathcal G)
\le
\Gamma_{\rm wit}(U;\mathcal G).
\end{equation}

When the certificate is optimized, its support may change. A confidence factor defined only from the
active coordinates would therefore depend on the candidate certificate. We instead use the certificate-independent factor
\begin{equation}
\kappa_m
=
\sqrt{2\ln(2m/\delta)},
\end{equation}
where $m$ is the size of the fixed full coordinate family. This choice gives a simultaneous confidence
bound that remains valid for every support encountered during the optimization.

In the ideal-centered, zero-bias continuous-budget model, the smallest budget required to make the
statistical penalty at most $\varepsilon$ is consequently
\begin{equation}
B_{\min}^{\rm phys}
=
\frac{
\kappa_m^2
\Gamma_{\rm phys}(U;\mathcal G)^2
}{
\varepsilon^2
}.
\end{equation}
Equation~\eqref{eq:joint-setting-cost} therefore gives an exact variational characterization of the
joint certificate-and-allocation problem for the fixed setting architecture $\mathcal G$.

The optimization is more difficult than the fixed-certificate allocation problem because
$\mathcal C_{\mathcal G}(y)$ is generally nonconvex over the feasible certificate set. This distinction
also limits what follows from the equal-stabilizer-norm separation. Theorem~\ref{thm:sep} proves that
two targets with the same stabilizer norm can have different physical robustness slopes
$s_\sdp$. It does not prove that the same targets have different values of
$\Gamma_{\rm phys}$. Establishing such an operational separation would require certified global bounds
for the nonconvex optimization in Eq.~\eqref{eq:joint-setting-cost}.

\subsection{Equal-shot benchmark}

When several observables share raw shots, allocation should be performed at the measurement-setting level. To isolate the effect of the robustness slope, suppose instead that every measured coordinate receives
the same number of shots. Define
\begin{equation}
q_\infty(z)
=
\max_i|z_i-b_i|,
\end{equation}
and let $\bar\beta$ be a uniform bound on the systematic error of each coordinate. The target-centered
certificate then gives
\begin{equation}
F_e
\ge
1-s\big[
q_\infty(z)+r_N+\bar\beta
\big].
\label{eq:acceptance}
\end{equation}

To certify
\begin{equation}
F_e\ge1-\varepsilon,
\end{equation}
it is sufficient that
\begin{equation}
q_\infty(z)+r_N+\bar\beta
\le
\frac{\varepsilon}{s}.
\end{equation}
If
\begin{equation}
q_\infty(z)+\bar\beta
<
\frac{\varepsilon}{s},
\end{equation}
the corresponding sufficient per-coordinate shot count is
\begin{equation}
N
\ge
\left\lceil
\frac{
2\ln(2m/\delta)
}{
\big[
\varepsilon/s
-q_\infty(z)
-\bar\beta
\big]^2
}
\right\rceil.
\label{eq:shots-general}
\end{equation}
The strict condition above is necessary for this bound to be useful: increasing the number of shots can
reduce statistical uncertainty, but it cannot compensate for an observed deviation and systematic
allowance that already exceed the acceptance threshold.

For ideal-centered data with no systematic bias,
\begin{equation}
q_\infty(z)=\bar\beta=0,
\end{equation}
Eq.~\eqref{eq:shots-general} reduces to
\begin{equation}
N
\ge
\left\lceil
\frac{2s^2}{\varepsilon^2}
\ln\frac{2m}{\delta}
\right\rceil.
\end{equation}
Before integer rounding, the physical and linear-witness benchmarks therefore satisfy
\begin{equation}
\frac{N_\sdp}{N_\wit}
=
\left(
\frac{s_\sdp}{s_\wit}
\right)^2.
\label{eq:shotratio}
\end{equation}

When
$s_\sdp<s_\wit$, the physical certificate requires fewer shots in this ideal-centered equal-shot
benchmark. The improvement is a constant-factor reduction rather than a change in the asymptotic
dependence on $\varepsilon$. For the three representative full-support two-qubit targets used in the
numerical study, the quantity
\begin{equation}
1-\left(\frac{s_\sdp}{s_\wit}\right)^2
\end{equation}
lies between $9.4\%$ and $11.4\%$.

These percentages illustrate the effect of the relaxation gap under the equal-shot benchmark; they are
not globally optimized experimental savings. Certification of empirical data must use
Eq.~\eqref{eq:acceptance}, and experiments with shared measurement settings are more appropriately
described by the grouped allocation problem of the preceding subsection.

Once a target-adapted dual certificate has been computed, applying it to new data requires only linear
post-processing. Computing that certificate requires an offline semidefinite optimization. Alternatively,
one may resolve the full physical SDP for each new empirical data set to obtain the tightest bound
within the chosen model.

For two qubits, the principal $240$-coordinate interface used to define the main robustness slopes is
channel-tomography-equivalent. The measurement-growth experiments below address a different question.
They remove the $15$ output-local marginals and select subsets of the $225$ two-register correlators,
producing genuinely incomplete measurement interfaces. Each such interface has its own exact-data and
finite-uncertainty certification problem and does not redefine the principal slopes studied above.

The comparison with direct fidelity estimation is therefore conceptual rather than a head-to-head
resource comparison. Direct fidelity estimation estimates the target fidelity using target-dependent
sampling of Pauli observables~\cite{FL,Leone,FawziOptimal}. The present formulation instead returns a
worst-case lower confidence bound over all physical channels consistent with the measured data. The
positivity constraint can strengthen that lower bound relative to the optimized linear-witness
relaxation.

\subsection{Sparse measurement design}

The measurement-family hierarchy determines what can be inferred from exact data, but it does not
determine how well the same family performs when the measurements are uncertain. A small set of
observables may uniquely identify the target at zero uncertainty and still give a poor fidelity
guarantee once finite statistical error is introduced. We examine this distinction numerically by
growing incomplete Pauli measurement families.

We performed $350$ exact-data semidefinite optimizations across five two-qubit targets. Starting from
subsets of the $225$ two-register correlators, we compared three ordering rules. The first adds
correlators in decreasing order of their ideal target magnitude. The second uses five fixed-seed random
orders. The third is an offline cutting-plane procedure: after solving the current worst-case
certification problem, it adds the eight unmeasured correlators on which the current adversarial state
differs most from the ideal target.

The cutting-plane rule uses the target and repeated optimization to design the measurement family before
data collection. It is therefore an offline target-specific selection rule, not a policy that adapts to
measurement outcomes during an experiment.

\begin{figure}[t]
\includegraphics[width=\columnwidth]{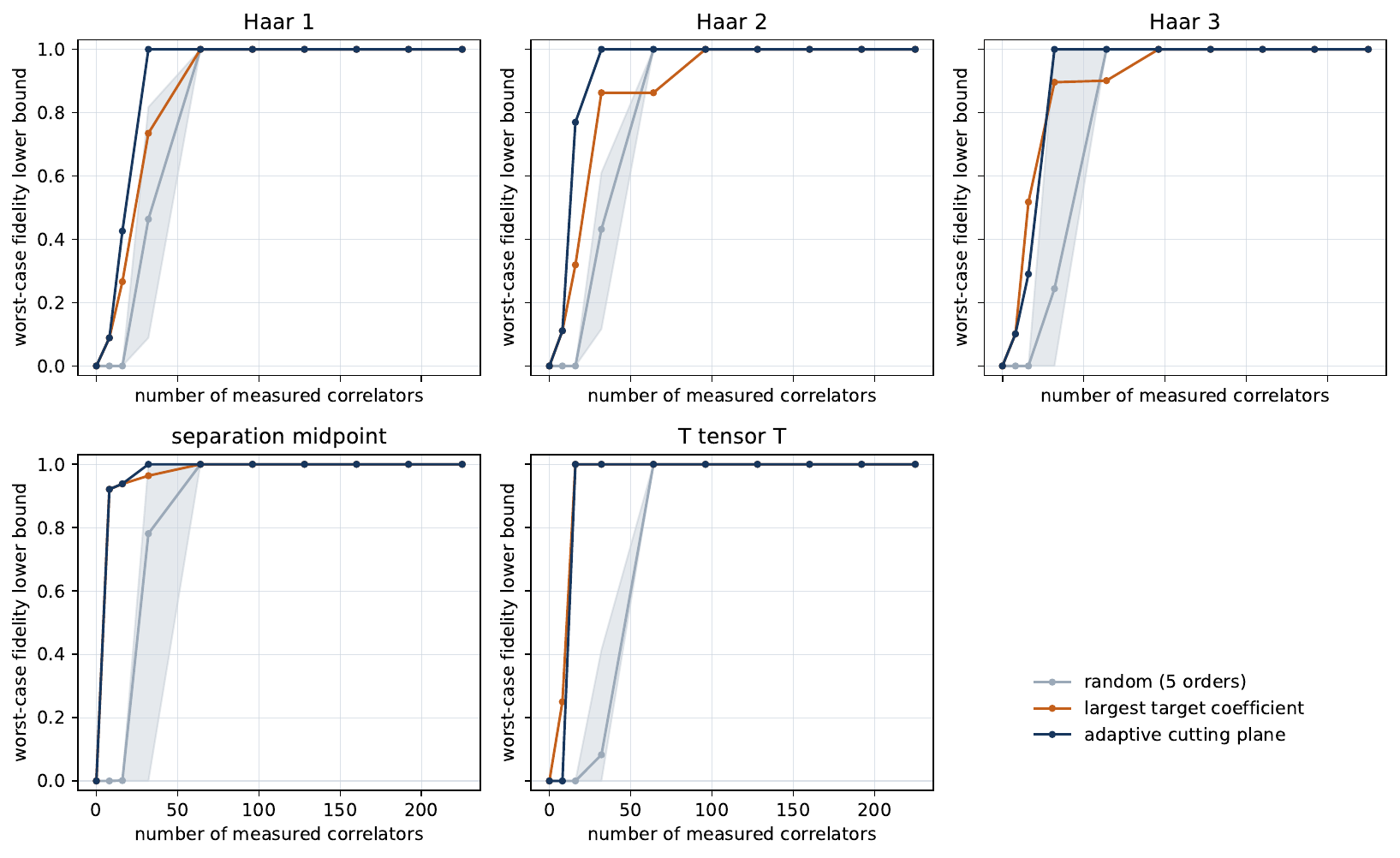}
\caption{Exact-data certification as the two-qubit measurement family is enlarged. Results are shown for
five targets and several correlator-selection rules. The offline cutting-plane procedure often raises
the worst-case fidelity close to one using substantially fewer than all $225$ two-register correlators.
The curves are numerical and do not establish global optimality of the selected subsets.}
\label{fig:growth}
\end{figure}

We then applied the same cutting-plane procedure to the archived sample of $200$ two-qubit Haar targets.
With $32$ selected correlators, repaired dual certificates gave an exact-data fidelity lower bound of at
least $0.99$ for $183$ targets. The remaining $17$ targets were extended to $64$ correlators, after
which every target had a bound above $0.99$. The corresponding selection logs, repaired certificates,
residuals, and solver statuses are reported in Appendix~\ref{app:numprot} and in the accompanying
archive.

These bounds assume exact expectation values. They do not imply that the selected sparse interfaces
are statistically efficient. The distinction is visible for the archived Haar-1 target. A
$32$-correlator subset obtained by ranking the full-interface dual weights still nearly identifies the
target at exact data, but after reoptimizing its finite-uncertainty certificate its robustness slope is
\begin{equation}
s=8.717.
\end{equation}
For the complete $225$-correlator family, the corresponding slope is only
\begin{equation}
s=2.630.
\end{equation}
Thus removing measurements can preserve near-perfect exact-data certification while making the
certificate substantially more sensitive to finite uncertainty.

For the same nested screening experiment, the smallest displayed equal-shot cost proxy occurs at
$64$ correlators. The subset with the lowest displayed equal-shot cost is therefore not the smallest subset tested. Adding measurements can improve the geometry of the
certificate, while measuring additional coordinates also consumes experimental resources.

\begin{figure}[t]
\includegraphics[width=\columnwidth]{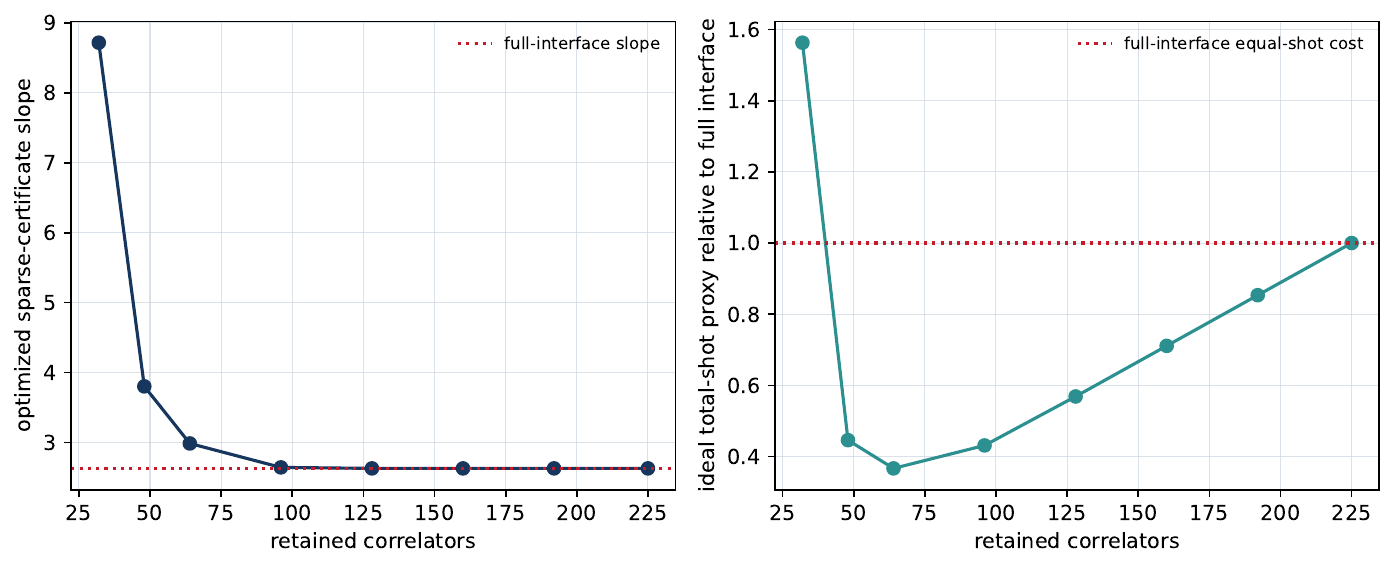}
\caption{Numerical sparsity-robustness tradeoff for the archived Haar-1 target. Interfaces that perform
well at exact data can have substantially different uncertainty slopes. Among the displayed nested
subsets, the equal-shot proxy is smallest at $64$ correlators. This comparison is numerical and does not
establish a globally optimal subset size.}
\label{fig:sparse-frontier}
\end{figure}

Theorem~\ref{thm:shot-allocation} solves one part of this design problem: once a certificate and a
coordinate family are fixed, it gives the optimal continuous allocation of coordinate shots under the
stated confidence model. It does not jointly select the observables, the certificate, and the compatible
measurement settings. The nested ranking used here is therefore a design heuristic rather than a
globally optimal measurement-selection procedure.

The Haar-1 example also illustrates the information hierarchy of
Theorem~\ref{thm:family-hierarchy}. For $d=4$, the real trace-preserving tangent space has dimension
\begin{equation}
d^4-d^2=240.
\end{equation}
The $225$ register-nontrivial Pauli correlators have rank $225$ on this space and leave a
$15$-dimensional output-local nullspace. Adding the $15$ output-local Pauli observables raises the rank
to $240$, making the full principal interface channel-tomography-equivalent.

The incomplete $225$-correlator family is nevertheless numerically consistent with target completeness
for Haar-1. Along the archived cutting-plane order, trace preservation alone gives numerical fidelity
lower bound zero; $16$ correlators give
\begin{equation}
L(0)=0.426064,
\end{equation}
$32$ give
\begin{equation}
L(0)=0.999996,
\end{equation}
and $64$ give
\begin{equation}
L(0)=0.999999994.
\end{equation}
These numerical values illustrate the progression from little fidelity information to near-unique target
identification. They are solver outputs and are not used as symbolic equality tests for target
completeness.

\subsection{Verification of the measurement interface}

The principal two-qubit interface can be implemented using compatible Pauli-basis measurements.
The resulting raw counts determine the expectation values required by the certificate.

The full interface contains $225$ two-register correlators together with $15$ output-local marginals.
All $240$ coordinates can be obtained from $81$ Pauli-basis settings. We implemented the complete map
from setting-level counts to the certification coordinates and then to the fidelity certificate. The
implementation was checked in both noiseless and finite-shot simulations, including setting-specific
calibration allowances.

The implementation was tested in simulation and has not been run on a physical quantum processor. These tests check that compatible measurement settings and raw counts produce the inputs required by the fidelity certificate. They also confirm the relevance of the
grouped allocation problem: several coordinates entering the dual certificate are estimated from the
same collection of raw shots. Converter tolerances, fixed-seed integration results, and the separate
simulated-noise stress test are reported in the reproducibility material.

As a small-instance check of the measurement-design analysis, we also exhaustively enumerated all
$512$ Pauli subsets for the one-qubit target $U_\star$. The enumeration and the corresponding figure are
reported in Appendix~\ref{app:numprot}.

\section{Discussion}
\label{sec:disc}

Our results distinguish linear Pauli robustness from robustness after enforcing positivity. For the linear relaxation, the optimized sensitivity is determined exactly by
the stabilizer norm,
\begin{equation}
s_\wit(U)=\|C_U\|_\st-\frac1D.
\end{equation}
After imposing positivity, the stabilizer norm alone no longer determines the robustness slope. Theorem~\ref{thm:sep} gives two distinct
two-qubit targets with exactly the same stabilizer norm but different positivity gains, and hence
different physical robustness slopes. The KKT characterization shows where the additional dependence
enters: through the support and signs of the Choi--Pauli coefficients together with the
target-dependent positivity constraint.

The spectral criteria detect positive gain without solving the full KKT completion problem. Product compression relates the gap to spectral properties of the Pauli sign
pattern, while the third moment identifies commuting Pauli multiplication triangles as a source of
sign-sensitive information. The eigenvalue and anticommuting-clique criteria give simpler sufficient
certificates for positive gain. These tests are not complete: the explicit clique-limitation example
shows that failure of a clique certificate does not settle the full KKT problem. Whether every
full-support two-qubit target has positive gain therefore remains open.

The dual coefficients also govern the effect of finite measurement resources on certification.
For a fixed certificate, its coefficients determine the optimal allocation of a shot budget, both for
individually sampled coordinates and for observables obtained from shared measurement settings.
Optimizing over the physical certificate leads to the setting-level cost
$\Gamma_{\rm phys}(U;\mathcal G)$. This is a stronger operational objective than the local slope
$s_\sdp$, and the equal-stabilizer-norm separation does not by itself imply a separation in
$\Gamma_{\rm phys}$.

Interval-certified calculations establish positive gain for the sampled finite target sets, while the larger
finite samples test the reach of the sufficient criteria. The sparse-measurement results show that a family can nearly identify the target from exact data yet remain sensitive to statistical uncertainty. We tested the full $81$-setting count-to-certificate pipeline in simulation; no hardware experiment is claimed.

The finite-data bounds assume trusted preparation and measurement and exact trace preservation. Our higher-dimensional results are limited to the finite sets of targets tested.

\section*{Data and code availability}

The code and data supporting the results of this study are publicly available in the
\href{https://github.com/shreyasavadatti/positivity-pauli-certification}{\texttt{positivity-pauli-certification}}
GitHub repository.

\begin{acknowledgments}
Grammarly was used for language editing.
\end{acknowledgments}

\clearpage
\appendix

\section{Supporting proofs}
\label{app:main-proofs}
This appendix collects the arguments supporting the statements in the main text.

\subsection{Proof of Proposition~\ref{prop:family}}
\begin{proof}
Every feasible $X$ has trace one. Since $C_U$ is a rank-one projector,
$\braket{C_U,X}=1$ holds if and only if $X=C_U$, proving the equivalence of the first two statements.
Writing $H=X-C_U$ proves the equivalence with the third statement. The displayed dual inequality gives
$\braket{C_U,X}\ge b^\top y+\Tr(\Lambda)/d=1$ for every feasible $X$, and hence is sufficient. Strict
dual feasibility follows from $y=0$ and $\Lambda=-c\Id_d$ with $c>0$, so the primal and dual values agree.
Therefore, if the equivalent conditions hold, feasible dual objectives approach one arbitrarily closely;
they reach one when the dual supremum is attained.
\end{proof}

\subsection{Proof of Theorem~\ref{thm:family-hierarchy}}
\begin{proof}
For physical $X$, $0\le\braket{C_U,X}\le1$. The first three statements follow directly from compactness
of the fiber, orthogonality at fidelity zero, and Proposition~\ref{prop:family} at fidelity one. If
$\ker\mathcal A\cap\mathcal V_{\rm TP}=\{0\}$, two trace-preserving Choi states with identical data have
zero difference and are equal. Conversely, let nonzero $H$ lie in this intersection and put
$X_0=\Id_D/D$, the full-rank depolarizing Choi state. For sufficiently small $\epsilon>0$, both
$X_0+\epsilon H$ and $X_0-\epsilon H$ are positive, trace preserving, distinct, and have identical data.
Thus the family is not tomography-equivalent.
\end{proof}

\subsection{Proof of Lemma~\ref{lem:tt}}
\begin{proof}
In the convention $\ket{\Phi_T}=(\ket{00}+e^{\mathrm i\pi/4}\ket{11})/\sqrt2$, define the one-qubit
$T$-Choi sign operator
\begin{equation}
\begin{aligned}
A_T&=Z\otimes Z+X\otimes X-Y\otimes Y\\
&\qquad+X\otimes Y+Y\otimes X\\[2pt]
&=\begin{pmatrix}1&0&0&2-2\mathrm i\\0&-1&0&0\\
0&0&-1&0\\2+2\mathrm i&0&0&1\end{pmatrix},
\end{aligned}
\end{equation}
whose spectrum is $\{1+2\sqrt2,\,1-2\sqrt2,\,-1,\,-1\}$, with
$A_T\ket{\Phi_T}=(1+2\sqrt2)\ket{\Phi_T}$. Choose the witness minimizer $\Lambda_0=\Id_d/D$, so that $T^*(\Lambda_0)=\Id_D/D$ removes the
global-identity coefficient. The resulting sign-fixed scaled subgradient can be written explicitly. Let $\mathsf P$ be the permutation taking the
pairwise Choi
ordering $(\mathrm{out}_1,\mathrm{ref}_1,\mathrm{out}_2,\mathrm{ref}_2)$ to the register ordering
$(\mathrm{out}_1,\mathrm{out}_2,\mathrm{ref}_1,\mathrm{ref}_2)$. Since the Choi coefficients and their
signs factorize for a product target, the bare scaled sign-fixed subgradient is
\begin{equation}
G_F=\mathsf P\big[(\Id_4+A_T)\otimes(\Id_4+A_T)-\Id_{16}\big]\mathsf P^\dagger,
\label{eq:GFfact}
\end{equation}
carrying the required signs on all $35$ nonzero register-nontrivial coefficients, with every free
scaled-subgradient coefficient set to zero. Every Pauli term in \eqref{eq:GFfact} is nontrivial on both Choi
registers, so $T(G_F)=0$.

The eigenvalues of $\Id_4+A_T$ are $2+2\sqrt2$, $2-2\sqrt2$, $0$, $0$, whence
\begin{equation}
\operatorname{spec}(G_F)=\big\{\,11+8\sqrt2,\ (-5)^{(2)},\ (-1)^{(12)},\ 11-8\sqrt2\,\big\},
\end{equation}
the eigenvalue $11+8\sqrt2$ belonging to
$\ket{\Phi_{T\otimes T}}=\mathsf P(\ket{\Phi_T}\otimes\ket{\Phi_T})$. All remaining eigenvalues are
strictly negative, since $11-8\sqrt2<0$ (because $121<128$). Hence with $\Pp=\Id-C_{T\otimes T}$ we have
$\Pp G_F\Pp\preceq0$, indeed strictly negative definite on $\ket{\Phi_{T\otimes T}}^\perp$. Both
conditions of Theorem~\ref{thm:kkt} hold, so $g(T\otimes T)=0$. The argument is exact algebra: no
eigensolver, interval computation or residual bound is involved.
\end{proof}

\subsection{Proof of Lemma~\ref{lem:t2}}
\begin{proof}
Expanding $R_\sigma(a)=\sum_{A,B}\sigma_{AB}\braket{a|A|a}B$ and using $\Tr(BB')=d\,\delta_{BB'}$,
\begin{equation*}
\Tr R_\sigma(a)^2=d\sum_B\sum_{A,A'}\sigma_{AB}\sigma_{A'B}\braket{a|A|a}\braket{a|A'|a}.
\end{equation*}
For traceless $A,A'$, the Haar pure-state moment identity~\cite{Harrow} gives
$\Eah\braket{a|A|a}\braket{a|A'|a}=\Tr(AA')/[d(d+1)]=\delta_{AA'}/(d+1)$, so
$\Eah\Tr R_\sigma(a)^2=\frac{d}{d+1}\sum_{A,B}\sigma_{AB}^2$. Since
$\Tr W_A^2=d\sum_B\sigma_{AB}^2$, this is $\frac1{d+1}\sum_A\Tr W_A^2$. For a full-support pattern all
$(d^2-1)^2$ coefficients are $\pm1$, giving
$\frac{d}{d+1}(d^2-1)^2=d(d^2-1)(d-1)$, independent of the signs.
\end{proof}

\subsection{Proof of Theorem~\ref{thm:anticlique}}
\begin{proof}
The columns of $E_{\mathcal C}$ are orthonormal and orthogonal to $\ket{\Phi_U}$. For any legal
completion $m\otimes\Id$, with $m=m^\dagger$ and $\Tr m=0$,
\begin{equation}
\operatorname{Re}\braket{e_A|m\otimes\Id|e_C}
=\frac{1}{2d}\Tr\!\big[m(CA+AC)\big].
\end{equation}
This vanishes for distinct $A,C\in\mathcal C$ by anticommutation and for $A=C$ by tracelessness of $m$.
Thus every real vector in the clique span has the same quadratic form for $G_F+m\otimes\Id$ as for
$G_F$. A positive eigenvector of $M_{\mathcal C}$ therefore contradicts the negative-semidefinite KKT
compression in Theorem~\ref{thm:kkt}. The stated six-clique incidence structure is the two-qubit maximum
anticommuting-set structure~\cite{Sarkar}. The embedded sum of the six clique compressions is exactly
$H$. If $x^{\mathsf T}Hx>0$, at least one clique contribution is positive, so the first part applies.
\end{proof}

\subsection{Proof of Proposition~\ref{prop:clique-limit}}
\begin{proof}
Appendix~\ref{app:clique-limit} gives a computer-assisted proof. Briefly, the stored binary64 matrix
$U_0$ is interpreted entrywise as an exact Gaussian-rational matrix and $Q$ is defined, without rounding,
as its unitary polar factor. Exact rational arithmetic gives a Pauli-support reserve greater than
$10^{-7}$ for $U_0$ and $\|U_0^\dagger U_0-\Id\|_F<10^{-13}$. The induced change in each Pauli expectation
is below $2.1\times10^{-13}$, so $Q$ has full support with the same signs. Exact Sylvester
tests make all six $-M_{\mathcal C}(U_0)$ positive definite. An inverse-norm bound gives a common lower
spectral gap greater than $0.1296$, while the polar-transfer change in each compression is less than
$10^{-9}$. Thus all six $M_{\mathcal C}(Q)$ remain negative definite.
\end{proof}

\subsection{Proof of Proposition~\ref{prop:onequbit}}
\begin{proof}
Write $R_{ij}(U)=\tfrac12\Tr(\sigma_iU\sigma_jU^\dagger)$ and $D_Y=\mathrm{diag}(1,-1,1)$. The transpose
on the reference Pauli basis gives $O_{ij}=R_{ij}(U)(D_Y)_{jj}$, i.e.\ $O=R(U)D_Y$; since
$R(U)\in SO(3)$ we obtain $O\in O(3)$ with $\det O=\det R\,\det D_Y=-1$.
The vectors $\ket{e_a}=(\sigma_a\otimes\Id)\ket{\Phi_U}$, $a\in\{x,y,z\}$, form an orthonormal basis of
$\ket{\Phi_U}^\perp$. Using
$\sigma_a\sigma_i\sigma_b=\delta_{ai}\sigma_b+\delta_{ib}\sigma_a-\delta_{ab}\sigma_i
+\mathrm i\epsilon_{aib}\Id$, the compression of $G_F=\sum_{ij}S_{ij}\sigma_i\otimes\sigma_j$ to this
basis is exactly $M(U)$. An admissible output-local completion $m\cdot\sigma\otimes\Id$ compresses to
$\mathrm iA(m)$ with $A(m)$ real antisymmetric, so $x^{\mathsf T}\mathrm iA(m)x=0$ for every real $x$.
If $M\preceq0$ the zero completion satisfies Theorem~\ref{thm:kkt}. Conversely, if $M$ has a positive
eigenvalue then a corresponding real eigenvector has positive expectation against every admissible
completion, so no completion makes the compression negative semidefinite. For $U_\star$ the displayed
matrices and eigenvalues follow by exact evaluation; $x=(1,-1,0)^{\mathsf T}$ gives
$x^{\mathsf T}M_\star x/x^{\mathsf T}x=\tfrac1{11}>0$.
\end{proof}

\section{Duality and the reduction to one observable}
\label{app:duality}
Strict dual feasibility holds: taking $y=0$ and $\Lambda=-c\Id_d$ with $c>0$ gives
$C_U-T^*\Lambda\succ0$, so strong duality applies to \eqref{eq:primal} by Slater's condition~\cite{Boyd}.
Dualizing \eqref{eq:primal} against $S\succeq0$ for positivity, $\Lambda$ for trace preservation, and $y$
for the data constraints, and using
$\min_{\|u\|_G\le\rho}\braket{y,u}=-\rho\|y\|_{G^\circ}$, gives
\begin{equation}
\begin{gathered}
L(z,\rho)=\max_{y,\Lambda}\Big[z^\top y+\tfrac1d\braket{\Id_d,\Lambda}
-\rho\|y\|_{G^\circ}\Big]\\
\text{s.t.}\quad C_U-\textstyle\sum_iy_iA_i-T^*\Lambda\succeq0.
\end{gathered}
\end{equation}
When $L_U(0)=1$, a dual certificate attaining this value satisfies
$b^\top y+\braket{\Id_d,\Lambda}/d=1$. Evaluating the same
feasible certificate at empirical center $z$ and radius $r$ gives
$L(z,r)\ge1+y^\top(z-b)-r\|y\|_{G^\circ}$, which is Eq.~\eqref{eq:empirical-affine} for the
$\ell_\infty$ box.
Strict dual feasibility ($y=0$, $\Lambda=-c\Id$) gives zero duality gap and primal attainment. For any
attained dual optimum at $\rho=0$ with $L(0)=1$, complementary slackness gives
$\braket{C_U,S}=0$, i.e.\ $S\ket{\Phi_U}=0$.
Finally, for the unrestricted measurement-design benchmark, the objective depends on the decomposition
only through $W=\sum_iy_iA_i$ and
$\min\{\|y\|_1:W=\sum_iy_iA_i,\ \|A_i\|_\op\le1\}=\|W\|_\op$, achieved by the single observable
$A=W/\|W\|_\op$. Thus the unrestricted optimum uses one designed observable. This reduction does not
apply to the fixed Pauli interface, where the available observables are prescribed coordinatewise.

\section{Proof of Theorem~\ref{thm:global}}
\label{app:global}
\emph{Upper bound.} Take $\Lambda=\tfrac12\Id_d$ and $S=0$; then
$\|C_U-\tfrac12\Id_D\|_\op=\max(1-\tfrac12,\tfrac12)=\tfrac12$ and $A^\star=2C_U-\Id_D$ has unit norm.
\emph{Lower bound.} For any feasible $(\Lambda,S)$ set $W=C_U-T^*\Lambda-S$ and choose a maximally
entangled $\ket\phi\perp\ket{\Phi_U}$, which exists for $d\ge2$: with the convention
$\ket{\Phi_U}=(U\otimes\Id)\ket{\Phi^+}$, take $\ket\phi=(UV\otimes\Id)\ket{\Phi^+}$ for a traceless
unitary $V$, so that $\braket{\Phi_U|\phi}=\tfrac1d\Tr V=0$. Both states have maximally mixed marginals, so
$\bra{\Phi_U}T^*\Lambda\ket{\Phi_U}=\bra{\phi}T^*\Lambda\ket{\phi}$. Using $S\ket{\Phi_U}=0$ and
$\bra{\phi}C_U\ket{\phi}=0$,
\begin{equation}
\bra{\Phi_U}W\ket{\Phi_U}-\bra{\phi}W\ket{\phi}=1+\bra{\phi}S\ket{\phi}\ \ge\ 1,
\end{equation}
whence $\|W\|_\op\ge\tfrac12$. \qed

\section{Proofs for the Pauli baseline}
\label{app:wit}
\begin{lemma}[Faithfulness]
\label{lem:faith}
Let $\rho_\psi=\ket\psi\!\bra\psi$ be a pure state on $2n$ qubits (dimension $D$).
Then $\|\rho_\psi\|_\st\ge1$, with equality iff $\ket\psi$ is a
stabilizer state; hence $\|C_U\|_\st-1\ge0$ with equality iff $U$ is Clifford.
\end{lemma}
\begin{proof}
By Parseval $\frac1D\sum_k|\Tr(P_k\rho_\psi)|^2=\Tr\rho_\psi^2=1$, and each $|\Tr(P_k\rho_\psi)|\le1$, so
$\frac1D\sum_k|\Tr(P_k\rho_\psi)|\ge\frac1D\sum_k|\Tr(P_k\rho_\psi)|^2=1$, with equality iff every coefficient lies
in $\{0,\pm1\}$. For a pure state, saturation implies that the nonzero unit-expectation Paulis form a
maximal stabilizer subgroup; hence $\ket\psi$ is a stabilizer state~\cite{Gottesman}. $\ket{\Phi_U}$ is a
stabilizer state iff $U$ is Clifford.
\end{proof}
\emph{Proof of Theorem~\ref{thm:wit}.} Since $C_U$ is maximally entangled its nonidentity Pauli support is
register-nontrivial, while $T^*\Lambda=\Id_d\otimes\Lambda$ has support only on labels trivial in the first
register. The output-local coordinates are included in the interface but vanish on $C_U$, while the
reference-local coordinates are eliminated by $\Lambda$. Any nonidentity reference-local component of
$\Lambda$ would add Pauli-$\ell_1$ cost, so an optimum may be chosen with no such component. Since the
global identity is omitted from $\|\cdot\|_{P,1}$, this gives
$\|C_U\|_{P,1}=\|C_U\|_\st-1/D$, which Lemma~\ref{lem:faith} splits as in \eqref{eq:floorsurcharge}. For
Clifford targets the support Paulis are signed stabilizers, $C_U=\frac1D\sum_{s\in\mathcal S}s$, and
$\widehat G=DC_U-\Id$ is a valid \emph{scaled} subgradient with $T(\widehat G)=0$ and
$\Pp\widehat G\Pp=-\Pp\preceq0$, so Theorem~\ref{thm:kkt}
gives tightness. \qed

\section{Proofs of Theorems~\ref{thm:kkt} and \ref{thm:dual}}
\label{app:kkt}
The proof is a first-order optimality argument. Variations in $\Lambda$ are unrestricted, whereas
the slack variations are positive semidefinite and supported on the target-orthogonal subspace.
The program $\min_{\Lambda,S}\{\|C_U-T^*\Lambda-S\|_{P,1}:S\succeq0,\ S\ket{\Phi_U}=0\}$ is convex. By the
normal-cone optimality condition~\cite{Boyd}, $(\Lambda_0,0)$ is optimal if and only if there exists a
true subgradient $G\in\partial\|\cdot\|_{P,1}(W_0)$ supporting every feasible direction
$-T^*\delta\Lambda-\delta S$. Writing $\widehat G=DG$, freeness of $\delta\Lambda$ gives
$T(\widehat G)=0$, while the cone of admissible $\delta S$ gives $\Pp\widehat G\Pp\preceq0$. Conversely,
if a scaled subgradient $\widehat G$ satisfies these two conditions then $G=\widehat G/D$ is a true
subgradient and
\begin{equation}
\|W\|_{P,1}\ \ge\ \|W_0\|_{P,1}+\tfrac1D\braket{\widehat G,W-W_0}\ \ge\ \|W_0\|_{P,1}
\end{equation}
for every feasible $W=C_U-T^*\Lambda-S$. For Theorem~\ref{thm:dual}, apply Sion's minimax theorem~\cite{Sion} to
the compact convex sets $\{g\in[-1,1]^{\mathcal R}\}$ and $\{Q\succeq0,\ Q=\Pp Q\Pp,\ \Tr Q=1\}$, using
$\min_{g}\braket{Q,\sum g_kP_k}=-\sum_k|\braket{Q,P_k}|$. \qed

\section{The equal-stabilizer-norm separation}
\label{app:sep}
The generator is the explicit traceless Hermitian matrix
\begin{equation}
\begin{gathered}
H=\frac14\begin{pmatrix}
3 & 1+\mathrm i & 0 & 2-\mathrm i\\
1-\mathrm i & -1 & \mathrm i & 1\\
0 & -\mathrm i & -2 & 1+2\mathrm i\\
2+\mathrm i & 1 & 1-2\mathrm i & 0
\end{pmatrix},\\[3pt]
\|H\|_F=\frac{\sqrt{42}}4,
\end{gathered}
\end{equation}
with $U(t)=e^{-\mathrm i tH}\in\SU(4)$ exactly, since $H^\dagger=H$ and $\Tr H=0$. For the proof, define
\begin{equation}
\begin{gathered}
f(t):=\|C_{U(t)}\|_\st-\|C_{T\otimes T}\|_\st,\\
t_0:=\frac{t_-+t_+}{2},\qquad w:=t_+-t_-.
\end{gathered}
\end{equation}
\paragraph*{Step 1: Equal-norm crossing.} Outward-rounded interval evaluation places $f(t_-)$ strictly below zero and
$f(t_+)$ strictly above zero, with the certified enclosures reported below. The normalized endpoint
enclosures are obtained by dividing the interval-certified difference of the raw $\ell_1$ sums of
$\widetilde c_k$ by $D=16$, as required by the definition of the stabilizer norm.
Continuity of $f$ therefore gives an exact equal-norm crossing in the bracket.

\paragraph*{Step 2: Sign stability.} Writing $\widetilde c_k(t)=\bra{\Phi(t)}P_k\ket{\Phi(t)}=D\,c_k(t)$ for the
unnormalized expectations, $\|\dot\Phi\|\le\|H\|_\op\le\|H\|_F$ and $\|P_k\|_\op=1$ give
$|\dot{\widetilde c}_k|\le2\|H\|_F$. The midpoint coefficient lower bound below exceeds
$\|H\|_F w$, so all 225 signs are constant throughout the bracket.

\paragraph*{Step 3: Positive-gain certificate.} With $F(t)=\Pp(t)G_F\Pp(t)$ and
$\|\dot C_U\|_\op\le2\|H\|_F$ one has $\|\dot F\|_\op\le4\|H\|_F\|G_F\|_F$, and
$|\lambda_{\max}(F(t))-\lambda_{\max}(F(s))|\le\|F(t)-F(s)\|_\op$ by Weyl. With the analytic bounds $\|G_F\|_F=60$ and $\|H\|_F=\sqrt{42}/4$ this
yields the numbers quoted in Theorem~\ref{thm:sep}. These bounds do not rely on a numerical eigensolver.
In particular, since $G_F$ is a $\pm1$ combination of
$|\mathrm{supp}\,G_F|=225$ orthogonal Pauli strings, $\|G_F\|_\op\le\|G_F\|_F=\sqrt{D\,|\mathrm{supp}|}=60$.
The eigenvalue lower bound uses a fixed trial vector $v$, a
nonzero \emph{rational} vector, obtained by rounding a numerical approximation to the midpoint top
eigenvector. Its rational coordinates and $H$ are stored in the accompanying data. Because the
scale-invariant quotient $\mathcal R_F(v)=v^\dagger F(t_0)v/(v^\dagger v)$ lower-bounds
$\lambda_{\max}(F(t_0))$ for \emph{any} nonzero $v$, only its direction matters, not its normalization,
and $v$ need be neither exactly normalized nor exactly an eigenvector. The certificate is summarized in
Table~\ref{tab:cert}. The Rayleigh bound minus the spectral drift exceeds $B_4$ throughout
the bracket, establishing positive gain at the crossing.

\paragraph*{Step 4: Interval implementation.}
All decisive scalars are enclosed by outward-rounded interval arithmetic
(\texttt{mpmath.iv}, \texttt{iv.dps=40}), with the Taylor remainder of $e^{-\mathrm i tH}$ added into
every entry as an interval. The implementation uses $48$ Taylor terms and, with
$x=|t|\sqrt{42}/4\ge |t|\|H\|_2$, bounds the omitted tail by $e^x x^{49}/49!$ in operator norm; the
corresponding symmetric rectangle is added to every matrix entry. The certified enclosures are
\begingroup\small
\begin{align}
f(t_-)&\in[-4.5690622787635,\,-4.5690622787633]\notag\\
&\qquad\times10^{-6}\subset(-\infty,0),\\
f(t_+)&\in[8.4074230946043,\,8.4074230946051]\notag\\
&\qquad\times10^{-6}\subset(0,\infty),\\
\textstyle\min_k|\widetilde c_k(t_0)|&\in[1.0343514961778,\,1.0343514961779]\notag\\
&\qquad\times10^{-5},\\
\mathcal R_F(v)&\in[18.366375587177,\,18.366375587178],
\end{align}
\endgroup
giving the certified reserve $\ge2.4324747912\times10^{-6}>0$ and the certified margin
$\mathcal R_F(v)-4\|H\|_F\|G_F\|_F w/2-B_4\in[11.65722232985,\,11.65722232986]$, strictly positive.
These strict inequalities establish the separation. Implementation details and independent
high-precision checks are given in Appendix~\ref{app:numprot} and the accompanying data.

\begin{table*}[t]
\caption{Certificate for Theorem~\ref{thm:sep}. Bracket
$[t_-,t_+]=[0.1758203125,\,0.1758251953]$, width $w=4.883\times10^{-6}$. All entries are either exact
analytic bounds or outward-rounded interval enclosures (\texttt{mpmath.iv}, \texttt{iv.dps=40}), with the
matrix-exponential truncation remainder included inside the intervals.}
\label{tab:cert}

\begin{tabular}{lc}
\toprule
requirement & certified enclosure or analytic bound \\
\midrule
$f(t_-)$ & $\le-4.5690622787633\times10^{-6}<0$\\
$f(t_+)$ & $\ge+8.4074230946043\times10^{-6}>0$\\
$\min_k|\widetilde c_k(t_0)|$ & $\ge1.0343514961778\times10^{-5}$\\
coefficient drift $2\|H\|_F\,w/2$ & $\le7.9110401705465\times10^{-6}$\\
sign-pattern reserve & $\ge2.4324747912\times10^{-6}>0$\\
Rayleigh quotient $\mathcal R_F(v)$ & $\ge18.366375587177$\\
$\|G_F\|_F$, $\|H\|_F$ (analytic) & $60$, $\sqrt{42}/4$\\
spectral drift $4\|H\|_F\|G_F\|_F\,w/2$ & $\le9.4932482047\times10^{-4}$\\
final margin over $B_4=\sqrt{45}$ & $\ge11.65722232985>0$\\
\bottomrule
\end{tabular}

\end{table*}

The construction involves no random seed: $H$ above is the complete specification.

\section{Product-compression certificate}
\label{app:prod}
Assume that $U$ has full register-nontrivial Pauli support, as in Theorem~\ref{thm:lp}.
After imposing $T(\widehat G)=0$, the free completion is therefore output-local.
For a unit vector $a$, the subspace $a\otimes(U^{\mathsf T}\ket{a^*})^\perp$ is orthogonal to $\ket{\Phi_U}$, and on it
a scaled subgradient $\Gs+m\otimes\Id$ compresses to $R_\sigma(a)+\braket{a|m|a}\Id$ restricted to a
$(d-1)$-dimensional subspace. If $g(U)=0$ then Theorem~\ref{thm:kkt} and Cauchy interlacing give
$\lambda_2(R_\sigma(a))\le-\braket{a|m|a}$ for every $a$; integrating against a measure with barycenter
$\Id_d/d$ and using $\Tr m=0$ yields $\int\lambda_2\,d\mu\le0$, which is the contrapositive of
\eqref{eq:lp}. For stabilizer-consistent patterns $R_\sigma(a)=d\,U^{\mathsf T}\ket{a^*}\!\bra{a^*}U^{*}-\Id$ has
$\lambda_2\equiv-1$, so $\int\lambda_2\,d\mu=-1$ for every probability measure $\mu$.

\subsection*{Proof of the eigenvalue budget (Theorem~\ref{thm:eig})}
Write the free completion as $m=\sum_{a\ne\Id}v_aP_a$ with $|v_a|\le1$. Pauli orthogonality gives
$\Tr m^2=d\sum_{a\ne\Id}v_a^2\le d(d^2-1)$. For a traceless Hermitian $m$ on $\mathbb C^d$ the extreme
eigenvalue obeys $\|m\|_\op^2\le\frac{d-1}{d}\Tr m^2$ (attained by the spectrum
$(\lambda,-\lambda/(d-1),\dots)$), so
$\|m\|_\op\le\sqrt{\tfrac{d-1}{d}\,d(d^2-1)}=\sqrt{(d^2-1)(d-1)}=B_d$. Weyl's inequality then gives
$\lambda_{\max}(\Pp(G_F+m\otimes\Id)\Pp)\ge\lambda_{\max}(\Pp G_F\Pp)-B_d$, which is positive whenever
$\lambda_{\max}(\Pp G_F\Pp)>B_d$; by Theorem~\ref{thm:kkt} no valid scaled subgradient exists and $g(U)>0$.

\section{Proof of Theorem~\ref{thm:skew}}
\label{app:skew}
Expanding $R_\sigma(a)=\sum_{A,B}\sigma_{AB}\braket{a|A|a}B$ gives
\begin{equation}
\begin{aligned}
\Tr R_\sigma(a)^3=\sum \sigma_{A_1B_1}\sigma_{A_2B_2}\sigma_{A_3B_3}&\\
\times\prod_{j}\braket{a|A_j|a}\ \Tr(B_1B_2B_3)&.
\end{aligned}
\end{equation}
For traceless $A_j$, the third Haar pure-state moment~\cite{Harrow} gives
$\Eah\prod_j\braket{a|A_j|a}=[\Tr(A_1A_2A_3)+\Tr(A_1A_3A_2)]/[d(d+1)(d+2)]$. Both $\Tr(A_1A_2A_3)$ and
$\Tr(B_1B_2B_3)$ vanish unless the third Pauli equals the product of the first two up to phase, which
produces multiplication triangles in both registers. If $A_1$ and $A_2$ anticommute, the two
first-register traces in the Haar moment cancel. Suppose instead that they commute. If $B_1$ and $B_2$
anticommute, pair each ordered summand with the summand obtained by exchanging the complete labels
$(A_1,B_1)$ and $(A_2,B_2)$. The product of the three $\sigma$ coefficients and the Haar factor are
unchanged, whereas $\Tr(B_2B_1B_3)=-\Tr(B_1B_2B_3)$, so the pair cancels. When both pairs commute, their
phases are real and the two first-register cycles add, giving the factor $2$. Terms with a repeated label
reduce to one-cycles and vanish by tracelessness, and terms with $\sigma_{AB}=0$ drop. Collecting the
prefactor $d\cdot 2/[d(d+1)(d+2)]\cdot d$ yields \eqref{eq:skew}. Invariance under the choice of Hermitian
representatives holds because a sign flip $A\mapsto-A$ flips $\sigma_{AB}$ and $\omega$ consistently around
each triangle. \qed

\section{Non-vanishing of the Choi-Pauli coefficients}
\label{app:nonzero}
Label a register-nontrivial string by an ordered pair $(P,Q)$ of nonidentity two-qubit Paulis, e.g.\
$(Z\otimes\Id,\,Z\otimes\Id)$. For the identity gate $C_{\Id}=\ket{\Phi^+}\bra{\Phi^+}$ one has
$c_{(P,Q)}(C_{\Id})=\Tr(P\otimes Q\,C_{\Id})/D=\Tr(PQ^{\mathsf T})/(dD)$, which is nonzero exactly when
$Q=P^{\mathsf T}$ up to phase; in particular $c_{(Z\otimes\Id,\,Z\otimes\Id)}(C_{\Id})\ne0$. Under
$U\mapsto C_1UC_2$ the coefficients obey
\begin{equation}
c_{P,Q}(C_1UC_2)=
c_{C_1^\dagger P C_1,\,C_2^*Q C_2^{\mathsf T}}(U).
\end{equation}
Thus the label pair is acted on independently in the two registers by the Clifford group, up to sign.
The Clifford group acts transitively on nonidentity Pauli strings~\cite{DehaeneDeMoor} (any such string is mapped to any other
by a suitable Clifford), so the product action is transitive on \emph{ordered pairs} of nonidentity
strings. Consequently every register-nontrivial label lies in the orbit of
$(Z\otimes\Id,\,Z\otimes\Id)$, and $c_k(C_1UC_2)=\pm c_{k'}(U)$ shows that no $c_k$ vanishes
identically. If a Clifford representative introduces a nonunit determinant, a compensating global phase
places it in $\SU(4)$ without changing the Choi projector or any Pauli coefficient. The union of their
zero sets is therefore a finite union of proper real-analytic zero sets
and has Haar measure zero~\cite{KrantzParks}.

\section{Exact certificate for the clique limitation}
\label{app:clique-limit}
The archived continuation data contain a $4\times4$ complex matrix $U_0$ at support floor $10^{-6}$.
Each binary64 real and imaginary part is interpreted by its exact
integer ratio, making $U_0$ an exact Gaussian-rational matrix. Define the exact target
$Q=U_0(U_0^\dagger U_0)^{-1/2}$.

Exact rational evaluation shows that all $225$
register-nontrivial Pauli expectations of $U_0$ are nonzero and the
smallest magnitude is $1.00000036690313849\times10^{-6}$, and that
$\|U_0^\dagger U_0-\Id\|_F^2<10^{-26}$. The exact rational residual is included in the archive.
The polar inequality
$\|Q-U_0\|_2\le\|U_0^\dagger U_0-\Id\|_2<10^{-13}$ therefore moves each unit-norm Pauli expectation by
less than $10^{-13}(2+10^{-13})$, which preserves every support sign.

For each of the six maximum cliques, exact positivity of all leading principal minors proves by
Sylvester's criterion that $-M_{\mathcal C}(U_0)$ is positive definite. Bounding
$\lambda_{\min}(A)\ge\|A^{-1}\|_\infty^{-1}$ for these six matrices gives respective lower bounds
$0.204254631643$, $0.219041755806$, $0.129650967987$, $0.182945205455$,
$0.358092446555$, and $0.162021741993$. Finally,
$\|G_F\|_2\le\|G_F\|_F=60$. Writing the perturbed tangent frame as $E=E_0+\delta E$ and expanding
$E^\dagger G_FE-E_0^\dagger G_FE_0$ into its two linear terms and one quadratic term gives
\begin{multline}
\|M_{\mathcal C}(Q)-M_{\mathcal C}(U_0)\|_2\\
\le60\sqrt5\,10^{-13}(2+\sqrt5\,10^{-13})<10^{-9}.
\end{multline}
The smallest exact gap exceeds this transfer allowance by more than eight orders of magnitude. Hence
all six $M_{\mathcal C}(Q)$ are negative definite while $Q$ retains full support. The raw rational
residual, all six inverse-norm bounds, and every pass/fail check are written to
the accompanying machine-readable JSON output.

\section{Proof of the optimal shot-allocation theorem}
\label{app:shot-allocation}

\begin{proof}[Proof of Theorem~\ref{thm:shot-allocation}]
Write $a_i=|y_i|>0$ for $i\in\mathcal A_y$. For an outcome bounded in $[-1,1]$, Hoeffding's inequality~\cite{Hoeffding}
gives
\begin{equation}
\Pr\!\left\{|z_i-\mathbb E z_i|>\frac{\kappa}{\sqrt{N_i}}\right\}
\le 2e^{-\kappa^2/2}=\frac{\delta}{k}.
\end{equation}
A union bound therefore gives simultaneous coverage of all active coordinates with probability at least
$1-\delta$. If the systematic displacement of coordinate $i$ is bounded in magnitude by $\beta_i$, the
fixed dual certificate implies on this event
\begin{equation}
F_e\ge 1+y^{\mathsf T}(z-b)
-\kappa\sum_{i\in\mathcal A_y}\frac{a_i}{\sqrt{N_i}}
-\sum_{i\in\mathcal A_y}a_i\beta_i.
\label{eq:allocation-proof-bound}
\end{equation}
It remains to minimize the statistical term.

The function $a_iN_i^{-1/2}$ is strictly convex on $N_i>0$, and it diverges as $N_i\downarrow0$.
Thus the budget-constrained problem has a unique interior minimizer. Its Lagrangian is
\begin{equation}
\mathcal L(N,\lambda)=\sum_{i\in\mathcal A_y}a_iN_i^{-1/2}
+\lambda\left(\sum_{i\in\mathcal A_y}N_i-B\right).
\end{equation}
Stationarity gives
\begin{equation}
-\frac{a_i}{2N_i^{3/2}}+\lambda=0,
\qquad
N_i=\left(\frac{a_i}{2\lambda}\right)^{2/3}.
\end{equation}
Let $A=\sum_{j\in\mathcal A_y}a_j^{2/3}$. Enforcing the budget gives
$(2\lambda)^{-2/3}=B/A$, and hence
\begin{equation}
N_i^\star=B\frac{a_i^{2/3}}{A}.
\end{equation}
Substitution yields
\begin{equation}
\sum_i\frac{a_i}{\sqrt{N_i^\star}}
=\frac{\sqrt A}{\sqrt B}\sum_i a_i^{2/3}
=\frac{A^{3/2}}{\sqrt B}
=\frac{\mathcal C_{2/3}(y)}{\sqrt B},
\end{equation}
which proves Eqs.~\eqref{eq:allocation-value} and~\eqref{eq:allocation-rule}. Inserting this value into
Eq.~\eqref{eq:allocation-proof-bound} proves Eq.~\eqref{eq:allocated-acceptance}.

For Corollary~\ref{cor:setting-allocation}, all coordinates in $\mathcal G_s$ use the same shot count,
so their contribution groups exactly as
\begin{equation}
\sum_{i\in\mathcal A_y}\frac{|y_i|}{\sqrt{N_{s(i)}}}
=\sum_{s=1}^{q}\frac{w_s(y)}{\sqrt{N_s}}.
\end{equation}
Applying the preceding strictly convex allocation argument with $a_i$ replaced by the positive setting
weights $w_s(y)$ gives Eq.~\eqref{eq:grouped-cost}. The probability statement still follows from the
coordinatewise tail bounds and a union bound; neither step assumes independence among estimators derived
from the same raw outcomes.

For completeness, suppose that $B\ge k$ is an integer and each $N_i$ must be a positive integer. After
assigning one shot to each active coordinate, define the reduction produced by the $(n+1)$st shot as
\begin{equation}
\Delta_i(n)=a_i\left(n^{-1/2}-(n+1)^{-1/2}\right),\qquad n\ge1.
\end{equation}
The sequence $\Delta_i(n)$ is nonincreasing because $x\mapsto x^{-1/2}$ is decreasing and convex.
Every integer allocation corresponds to selecting, for each coordinate, an initial segment of these
marginal reductions. Maximizing the total reduction from the one-shot baseline is therefore achieved by
selecting the largest available marginal reduction at each of the remaining $B-k$ steps. This proves the
stated greedy rule for the exact integer optimum.
\end{proof}

\section{Numerical methods and reproducibility}
\label{app:numprot}
Haar unitaries are generated by the phase-corrected QR construction for complex Ginibre matrices~\cite{Mezzadri}, with fixed seeds. Semidefinite
programs are solved with CLARABEL~\cite{Clarabel} and cross-checked with SCS~\cite{SCS} through CVXPY~\cite{CVXPY}; solver versions, termination
statuses and primal/dual residuals accompany the reproducibility package. Support is classified at
threshold $10^{-10}$. The equal-stabilizer-norm target is obtained by the validated bracket procedure of Theorem~\ref{thm:sep}, not by bisection to a tolerance. For the one-qubit prevalence experiment, $2\times10^5$ Haar-random $\SU(2)$
unitaries were generated by QR decomposition of complex Ginibre matrices with phase correction, followed
by division by $\sqrt{\det}$, using \texttt{numpy} \texttt{default\_rng} seed $7$; full support was
classified at threshold $10^{-12}$ on $|O_{ij}|$, and positive gain was recorded when
$\lambda_{\max}(M)>10^{-12}$. The quoted binomial standard error is computed from the observed success
fraction. Monte Carlo moment estimates are reported with standard errors; the triangle statistic $\Delta_c$ is enumerated exactly, so structured
controls (CNOT, $T\otimes T$) are exact analytic checks while Haar-random targets are compared within sampling
error. An independent solver-based cross-check, \emph{not used in Theorem~\ref{thm:sep}}, computes the dual
obstruction $Q$ of Theorem~\ref{thm:dual} and confirms the same conclusion. Quantities described as
``certified'' are those for which the certificate remains valid after
accounting for numerical error; quantities obtained only as solver optima are labeled as such.

\subsection{Additional numerical results}

For the $200$ two-qubit and ten $d=8$ interval-certified targets, every reported certificate value is a
Rayleigh-quotient lower bound $\mathcal R_F(v)$ evaluated on a stored trial vector with a conservative
absolute allowance of $10^{-9}$; each stored binary64 matrix defines an exact target as its polar unitary.
The outward-rounded interval calculation interprets the stored entries and rational trial vectors exactly, evaluates
every Rayleigh quotient in interval arithmetic, and transfers it to the polar unitary using the
interval-bounded unitarity residual. The largest spectral-transfer bound is $2.07\times10^{-13}$ at $d=4$
and $1.89\times10^{-12}$ at $d=8$, both below the allowance, and the smallest sign reserves after transfer
are $1.11\times10^{-6}$ ($d=4$) and $9.90\times10^{-7}$ ($d=8$).

In the ten-seed clique ablation, every distinct dimension-one through dimension-five subset of the six
maximum cliques gave positive-certificate counts $1$, $9{,}996$, $10{,}000$, $10{,}000$, and $10{,}000$
over $10{,}000$ targets. The dimension-two success rate ranged from $99.8\%$ to $100\%$ across seed
blocks, while dimensions three through five were $100\%$ in every block; for dimension three the per-seed
minimum certificate value had mean $0.98699$ and standard deviation $0.12274$.

In the sparse-design study, the median $32$-correlator bound was $0.999999977$ and the hardest target gave
$0.545704$; after extension to $64$ correlators every repaired lower bound exceeded $0.999999974$, with
largest primal-minus-dual gap $1.41\times10^{-8}$. At $64$ correlators the equal-shot proxy slope was
$2.986$ and the proxy cost $0.367$ of the $225$-correlator value. Complete weights, trace multipliers,
slack repairs, and selection logs are archived.

Table~\ref{tab:num-summary} records the sample sizes, quantitative ranges, and evidentiary status used in
the numerical section. It is collected here so that the main text can focus on the scientific
conclusions.

\begin{table*}[t]
\caption{Summary of numerical calculations and evidentiary status. $B_4=\sqrt{45}$ and $B_8=21$.}
\label{tab:num-summary}
\small
\begin{tabular}{lccccc}
\toprule
experiment & $d$ & samples & quantity & range & status\\
\midrule
KKT vs.\ direct SDP & $4$ & $5$ & sign agreement & $5/5$ & solver-observed\\
Eigenvalue certificate & $4$ & $200$ & margin over $B_4$ & $\ge0.773$ & interval-certified\\
Anticommuting cliques & $4$ & $10{,}000$ & max clique eigenvalue & $\ge1.0038$ & numerical\\
Clique ablation & $4$ & $10\times1{,}000$ & dimension-three success & $10{,}000/10{,}000$ & numerical\\
Clique limitation & $4$ & $1$ & gap of $-M_{\mathcal C}$ & $\ge0.12965$ & exact/polar cert.\\
Adversarial full KKT & $4$ & $1$ path & minimum completion bound & $5.545$ & numerical\\
Eigenvalue certificate & $8$ & $10$ & margin over $B_8$ & $\ge37.38$ & interval-certified\\
Triangle formula & $4$ & exact $+$ $600$ Monte Carlo & $\Delta_c$ vs.\ Monte Carlo & $<1\%$ & statistical\\
Separation certificate & $4$ & $1$ & margin over $B_4$ & $>11.65$ & interval-certified\\
One-qubit criterion & $2$ & $2{\times}10^5$ & $M(U)\not\preceq0$ & $1.504\%$ & statistical\\
\bottomrule
\end{tabular}
\end{table*}

The complete KKT search used $240$ random perturbations of $T\otimes T$ at six scales from $10^{-6}$
to $5\times10^{-2}$ and $60$ Haar controls. The smallest evaluated completion value was $5.518$ in the
Haar group and $8.225$ near $T\otimes T$. In the separate adversarial continuation, the smallest observed
complete KKT value was $5.54519$ while the support floor was continued through $10^{-8}$, $10^{-9}$, and
$10^{-10}$. These nonconvex binary64 searches are evidence, not certified global minima.

For the three-qubit extension, all $200$ stored Haar targets satisfied
$\lambda_{\max}(\Pp G_F\Pp)>B_8=21$. The minimum margin was $34.498$, the median was $39.459$, and the
largest eigenvector residual was $1.16\times10^{-13}$. The separate ten-target interval-certified subset
provides rigorous certificates and certified open neighborhoods.

Finally, all $2^9=512$ Pauli subsets were enumerated for the positive-gain one-qubit target $U_\star$.
Exactly $226$ subsets returned a finite robustness certificate, and none with fewer than four
correlators was feasible. The best equal-shot proxy uses
$\{X\otimes Y,Y\otimes X,Z\otimes X,Z\otimes Y\}$, has slope $1.2580645$, and costs $0.737604$ relative
to the nine-correlator interface. A seven-correlator subset matches the full-interface slope to
$6\times10^{-9}$ relative accuracy. Exhaustiveness refers only to this one-qubit instance and the stated
solver tolerances.

\begin{figure}[t]
\includegraphics[width=\columnwidth]{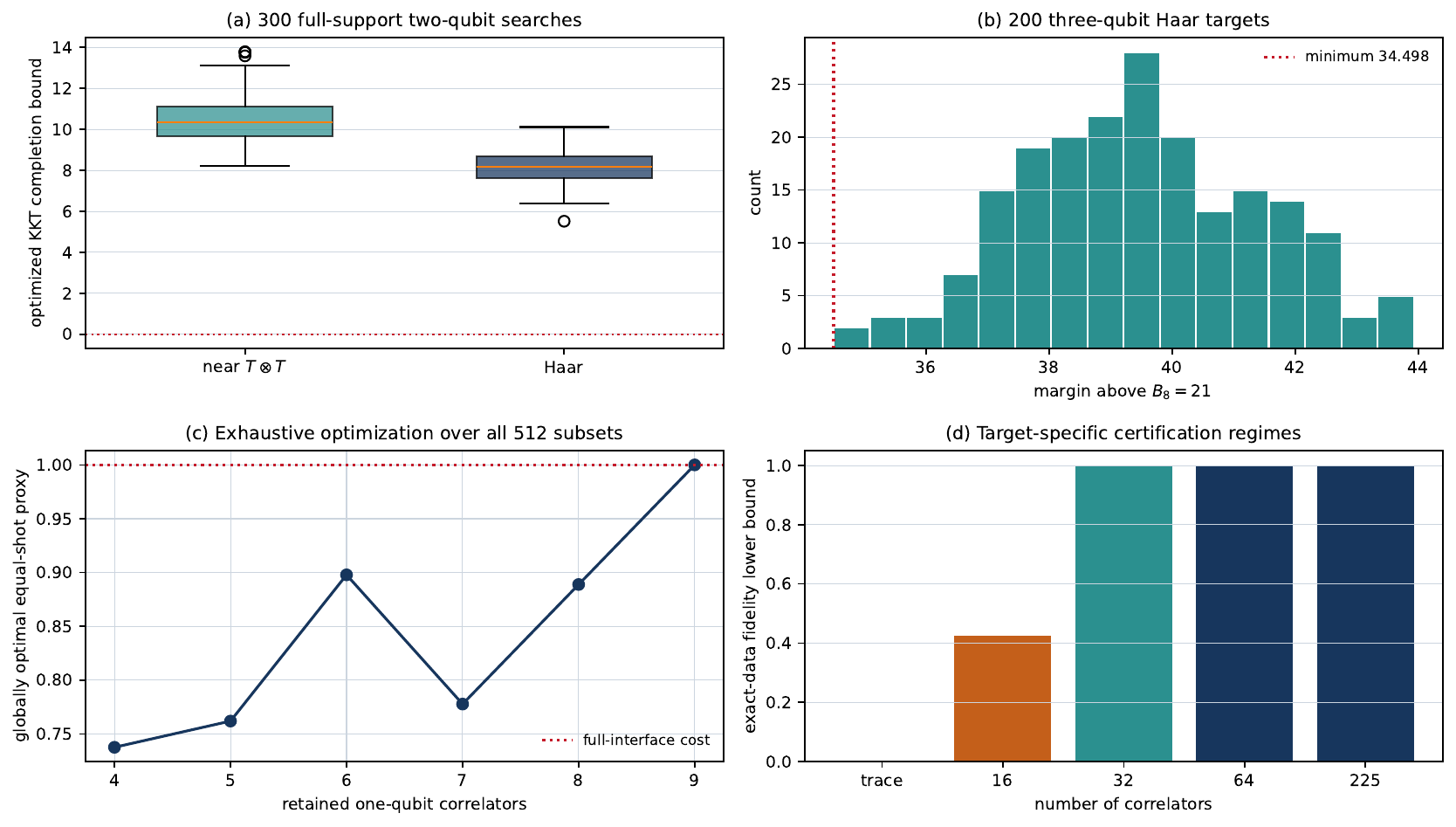}
\caption{Additional numerical results. (a) The KKT search finds no full-support two-qubit zero-gain candidate.
(b) All $200$ stored three-qubit targets numerically exceed the sufficient budget. (c) Exhaustive
one-qubit subset optimization gives a small-instance benchmark. (d) The archived Haar-1 target moves
from uninformative to partial and then numerically target-complete as correlators are added.}
\label{fig:extensions}
\end{figure}

\end{document}